\documentclass[12pt]{article}

\usepackage{graphicx}
\usepackage{natbib}
\usepackage[table,usenames,dvipsnames]{xcolor}
\usepackage[margin=1.5cm]{geometry}
\usepackage[hypertexnames=false,colorlinks=true,breaklinks]{hyperref}
\usepackage{rotating}
\usepackage{verbatim}
\usepackage{caption}
\usepackage{url}
\usepackage[toc,nonumberlist,numberedsection=autolabel]{glossaries}
\usepackage{endfloat}
\usepackage{siunitx}
\usepackage{amsmath}

\newglossaryentry{PAC} {
    name={PAC},
    description={phase-amplitude coupling. Coupling between the phase of
low-frequency oscillations and the amplitude of high-frequency ones}
}

\newglossaryentry{ERP} {
    name={ERP},
    description={event-related potential. Mean of voltages recorded at a single
electrode and aligned with respect to an event of interest (e.g., the
initiation of the production of \glspl{CVS}}
}

\newglossaryentry{MI} {
    name={MI},
    description={modulation index quantifying the strength of phase-amplitude coupling}
}

\newglossaryentry{TW} {
    name={TW},
    description={traveling wave}
}

\newglossaryentry{CVS} {
    name={CVS},
    description={consonant vowel syllable}
}

\newglossaryentry{vSMC} {
    name={vSMC},
    description={ventral sensorimotor cortex. A brain region that controls the
vocal articulators}
}

\newglossaryentry{ECoG} {
    name={ECoG},
    description={electrocorticography. A brain recording modality that measures
potentials directly from the cortical surface}
}

\newglossaryentry{MEG} {
    name={MEG},
    description={magnetoencephalography. A brain recording modality that 
measures magnetic fields from the scalp}
}

\newglossaryentry{EEG} {
    name={EEG},
    description={electroencephalography. A brain recording modality that 
measures electrical potentials from the scalp}
}

\newglossaryentry{PLI} {
    name={PLI},
    description={phase-locking index. A measure of alignment among multiple phases}
}

\newglossaryentry{ERSP} {
    name={ERSP},
    description={event-related spectral perturbation. An average measure of evoked power across trials}
}

\newglossaryentry{cvsPhase} {
    name={CVS phase},
    description={phase of filtered-voltage oscillation at which a \gls{CVS} is
initiated}
}

\makeglossaries

\begin{document}


\title{Rhythmic production of consonant-vowel syllables synchronizes traveling
waves in speech-processing brain regions}

\author{Joaqu\'{i}n Rapela\thanks{rapela@ucsd.edu}}
\maketitle

\abstract{
Nature is abundant in oscillatory activity, with oscillators that have the
remarkable ability of synchronizing to external events.
Using electrocorticographic (\gls{ECoG}) recordings from a subject rhythmically
producing consonant-vowel syllables (\glspl{CVS}) we show that neural
oscillators recorded at individual \gls{ECoG} electrodes become precisely
synchronized to initiations of the production of \glspl{CVS} (i.e., that these
initiations occur at precise phases of bandpassed-filtered voltages recorded at
most \gls{ECoG} electrodes).
This synchronization is not a trivial consequence of the rhythmic production of
\glspl{CVS}, since it takes several minutes to be fully established and is
observed at the frequency of \gls{CVS} production and at its second harmonic.
The phase of filtered voltages at which \glspl{CVS} are produced varies
systematically across the grid of electrodes, consistently with the propagation
of traveling waves (\glspl{TW}).
Using these synchronized phases we isolate a first \gls{TW} in voltages
(filtered at the median \gls{CVS}-production frequency) moving from primary
auditory to premotor cortex, and a second \gls{TW} in high-gamma amplitude
(coupled to phase at the \gls{CVS}-production frequency) moving along the same
path but in opposite direction.
To our knowledge, this is the first report of rhythmic motor acts synchronizing
spatio-temporally organized cortical activity in the human brain.

}

\pagebreak

\tableofcontents

\pagebreak
\setcounter{page}{1}

\section{Introduction}

Physiological function emerges from interactions of neurons with each other and
with external inputs to generate rhythms essential to life~\citep{glass01}.
The heartbeat results from interactions of thousand of pacemaker cells in the
right atrium of the heart~\citep{defeliceAndIsaac93,guevaraEtAl95}.
Nerve cells generating locomotion respond with precise phase relation depending
on the type of gait~\citep{golubitskyEtAl99}.
And the sleep-wake rhythm is usually synchronized to the light-dark
cycle~\citep{winfree01,strogatz86}.
Physiological oscillations can be synchronized to appropriate external or
internal stimuli. Plants and animals show a circadian rhythm in which key
processes show a 24-hour periodicity, which is usually set by the 24-hour
light-dark cycle. Thus, the dark-light cycle synchronizes the intrinsic
rhythm of plants and animals~\citep{winfree01,strogatz86}.
Periodic inputs from medical devices can synchronize functions of the body. For
example, a mechanical ventilator delivers mixes of gases to patients in a
periodic fashion. The resulting lung inflation interacts with the patient's
intrinsic respiratory rhythms so that the patient respiratory rhythm is
entrained to the ventilator~\citep{petrilloAndGlass84,gravesEtAl86,simonEtAl00}.

External visual~\citep{regan66} and auditory~\citep{galambosEtAl81} rhythmic
stimuli entrain brain oscillations (i.e., drag brain oscillations to follow the
rhythm of the stimuli), and this entrainment is modulated by attention so that
the occurrence of attended stimuli coincides with the phase of brain
oscillations of maximal
excitability~\citep{lakatosEtAl05,lakatosEtAl08,lakatosEtAl13,oconnellEtAl11,besleEtAl11,gomezRamirezEtAl11,zionGolumbicEtAl13,cravoEtAl13,mathewsonEtAl10,spaakEtAl14,grayEtAl15}.
Although speech is only quasi rhythmic~\citep{cummins12} an increasing number of
studies is showing that neural oscillations can entrain to speech
sound~\citep[e.g.,][]{zionGolumbicEtAl13b,grossEtAl13,doellingEtAl14,millmanEtAl15,parkEtAl15}.
For recent reviews on the entrainment of neural oscillations to speech sound
see \citet{peelleAndDavis12,zionGolumbicEtAl12,dingAndSimon14}.

What is currently unknown is whether rhythmic motor acts can synchronize neural
oscillators. In a previous article~\citep{rapelaInPrepTWsInSpeech} we reported that rhythms in
electrocorticographic (\gls{ECoG}) recordings from speech processing brain
regions in the left hemisphere of a subject rhythmically producing
consonant-vowel syllables (\glspl{CVS}) are spatio-temporally organized as
traveling waves (\glspl{TW}). We also showed that the coupling between phase
at the \gls{CVS}-production frequency and amplitude in the high-gamma range was
spatially organized in agreement with the propagation of \glspl{TW}. Here we
demonstrate that the
previously reported \glspl{TW} become precisely synchronized to the production
of \glspl{CVS}.

The synchronization of neural oscillators to rhythmic inputs can (1)
be precise, (2) take time to develop, and (3) operate on oscillators at
different temporal scales~\citep[Arnold tongues;][Chapter 10]{izhikevich07}.
To motivate our findings in \gls{ECoG} recordings, in
Section~\ref{sec:syncInNeuralModels} we illustrate these synchronization
concepts in models of single-neuron oscillators.
In
Section~\ref{sec:syncInNeuralPopulations} we show that the synchronization of
neural oscillators over speech production regions (1) is precise, (2) takes
time to develop, and (3) operates on oscillators at the median frequency of
\gls{CVS} production and at its second harmonic.
Section~\ref{sec:cvsPhaseaAcrossElectrodes} shows that the phases of filtered
voltages at which \glspl{CVS} are initiated (i.e., the \glspl{cvsPhase}) vary
systematically across the grid of \gls{ECoG} electrodes, consistently with the
propagation of \glspl{TW}. Using these synchronized phases in
Section~\ref{sec:a1PremotorTWs} we isolate a first \gls{TW} synchronized to the
production of \glspl{CVS} in voltages filtered around the median
\gls{CVS}-production frequency and moving from premotor to auditory
cortex, and a second \gls{TW}, also synchronized to the production of
\gls{CVS}, but in high-gamma amplitude coupled with phase at the
median \gls{CVS}-production frequency, and traveling along the same path but in
opposite direction as the previous \gls{TW}.

\section{Results}

\subsection{Synchronization of models of single-neuron oscillators to periodic
input pulses}
\label{sec:syncInNeuralModels}

In this section we illustrate how simulated oscillators receiving periodic
inputs (stimulation frequency $f_s=0.15$~Hz) can synchronize to these inputs.
We first demonstrate a 1-to-1 synchronization (i.e., one oscillator cycle
paired to one input) of an oscillator with a natural oscillation frequency
close to the stimulation frequency (Section~\ref{sec:1-1SyncModels}).
Next we exemplify a 2-to-1 synchronization (two oscillator cycles paired to
one input) of an oscillator with a natural frequency at the second harmonic of
the stimulation frequency (Section~\ref{sec:2-1SyncModels}).

\subsubsection{1-to-1 synchronization at the fundamental frequency}
\label{sec:1-1SyncModels}

\begin{figure}
\begin{center}
\includegraphics[width=4in]{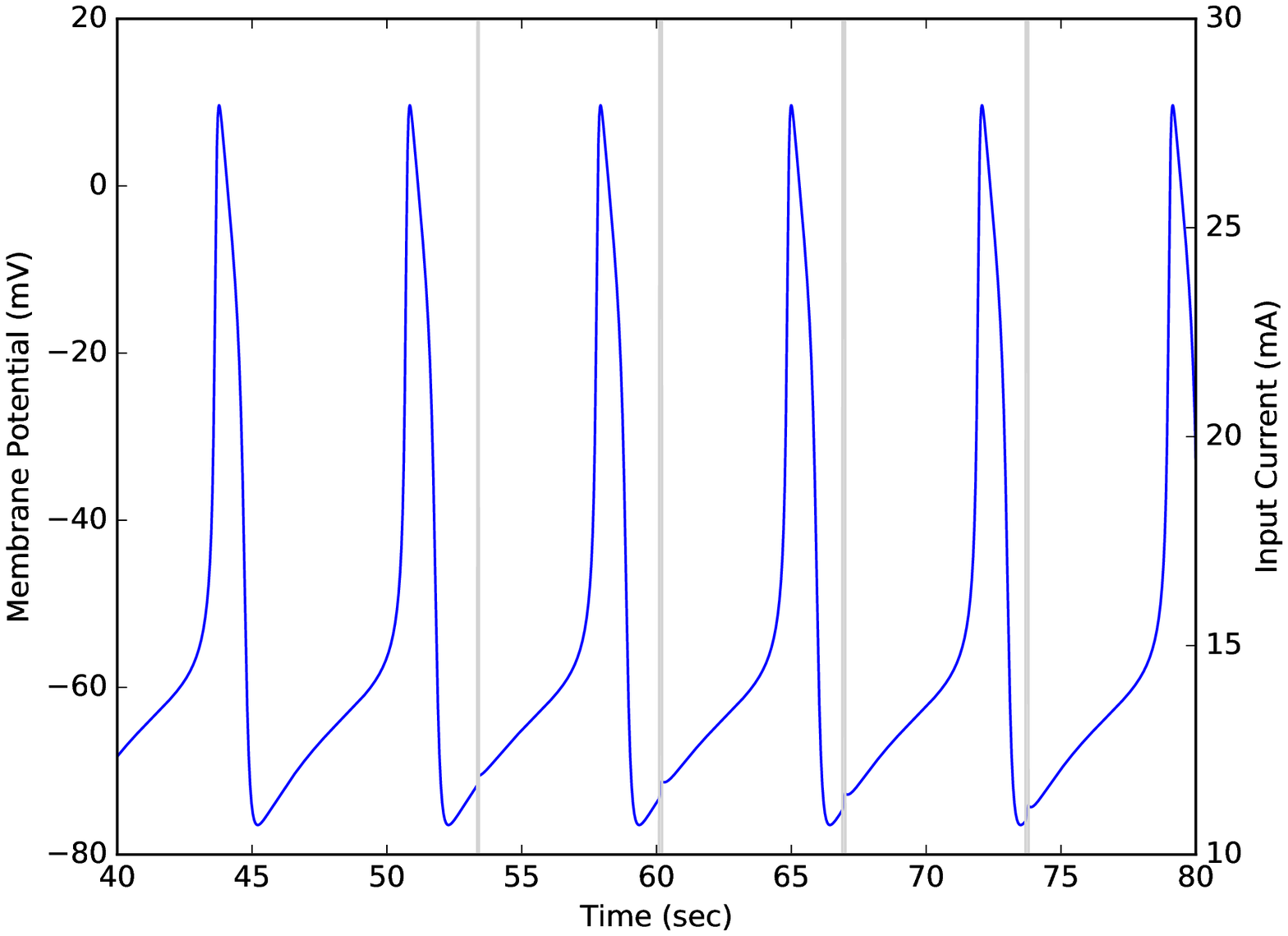}
\end{center}

\caption{Lack of synchronization of a single-neuron model to periodic rhythmic
stimulation at the beginning of the stimulation period. Before stimulation the
neural model was running on a limit cycle with frequency close to that of
stimulation.}

\label{fig:1-1SyncNeuralModelBegining}
\end{figure}

\begin{figure}
\begin{center}
\includegraphics[width=4in]{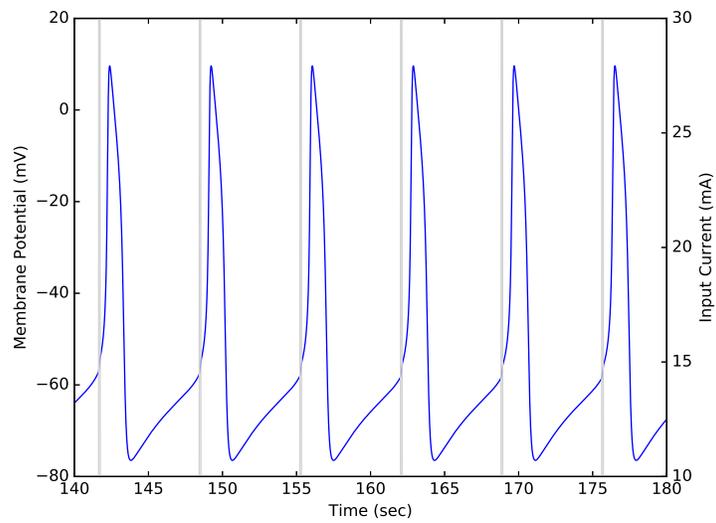}
\end{center}

\caption{Precise synchronization of a single-neuron model to periodic rhythmic
stimulation after 160~seconds of stimulation. Before stimulation the
neural model was running on a limit cycle with frequency close to that of
stimulation.}

\label{fig:1-1SyncNeuralModelLater}
\end{figure}

\begin{figure}
\begin{center}
\includegraphics[width=4in]{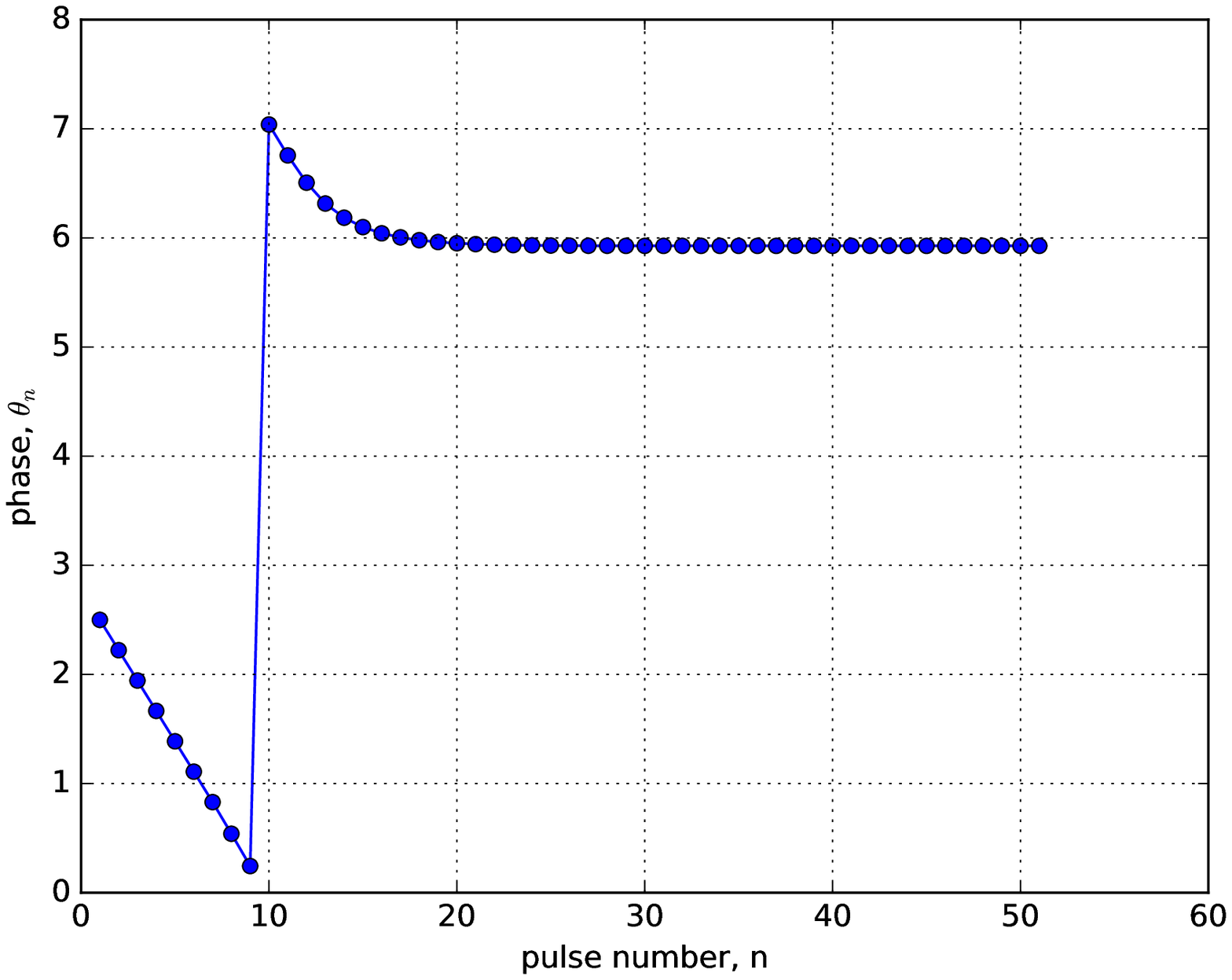}
\end{center}

\caption{Temporal evolution of synchronization across time.
Blue points indicate the phase of the spiking cycle at which current inputs
were delivered. Before stimulation the neural model was running on a limit cycle
with frequency close to that of stimulation.}

\label{fig:1-1SyncNeuralModelAllTimes}
\end{figure}

Figures~\ref{fig:1-1SyncNeuralModelBegining}
and~\ref{fig:1-1SyncNeuralModelLater} show the membrane potential of an
oscillator (blue trace) with natural frequency close to the stimulation
frequency (see Slower Oscillator in Section~\ref{sec:oscillatorModels}) and its
periodic current pulses inputs (gray vertical lines with scale on the right
axis) at the beginning of the stimulation period and after 150~seconds of
stimulation, respectively. At the beginning of the stimulation the oscillator is
not synchronized to its inputs (Figure~\ref{fig:1-1SyncNeuralModelBegining}),
but after 150~seconds it is already synchronized
(Figure~\ref{fig:1-1SyncNeuralModelLater}).
Figure~\ref{fig:1-1SyncNeuralModelLater} illustrates 1-to-1 synchronization in
an oscillator whose natural oscillation frequency is close to the stimulation
frequency.
Figure~\ref{fig:1-1SyncNeuralModelAllTimes} plots phases of the oscillatory
cycle at which input pulses arrives (i.e., \glspl{cvsPhase}). After 20 input
pulses (136~seconds) these phases become constant, indicating that the
oscillator is synchronized to its inputs.

\subsubsection{2-to-1 synchronization at the second harmonic}
\label{sec:2-1SyncModels}

\begin{figure}
\begin{center}
\includegraphics[width=4in]{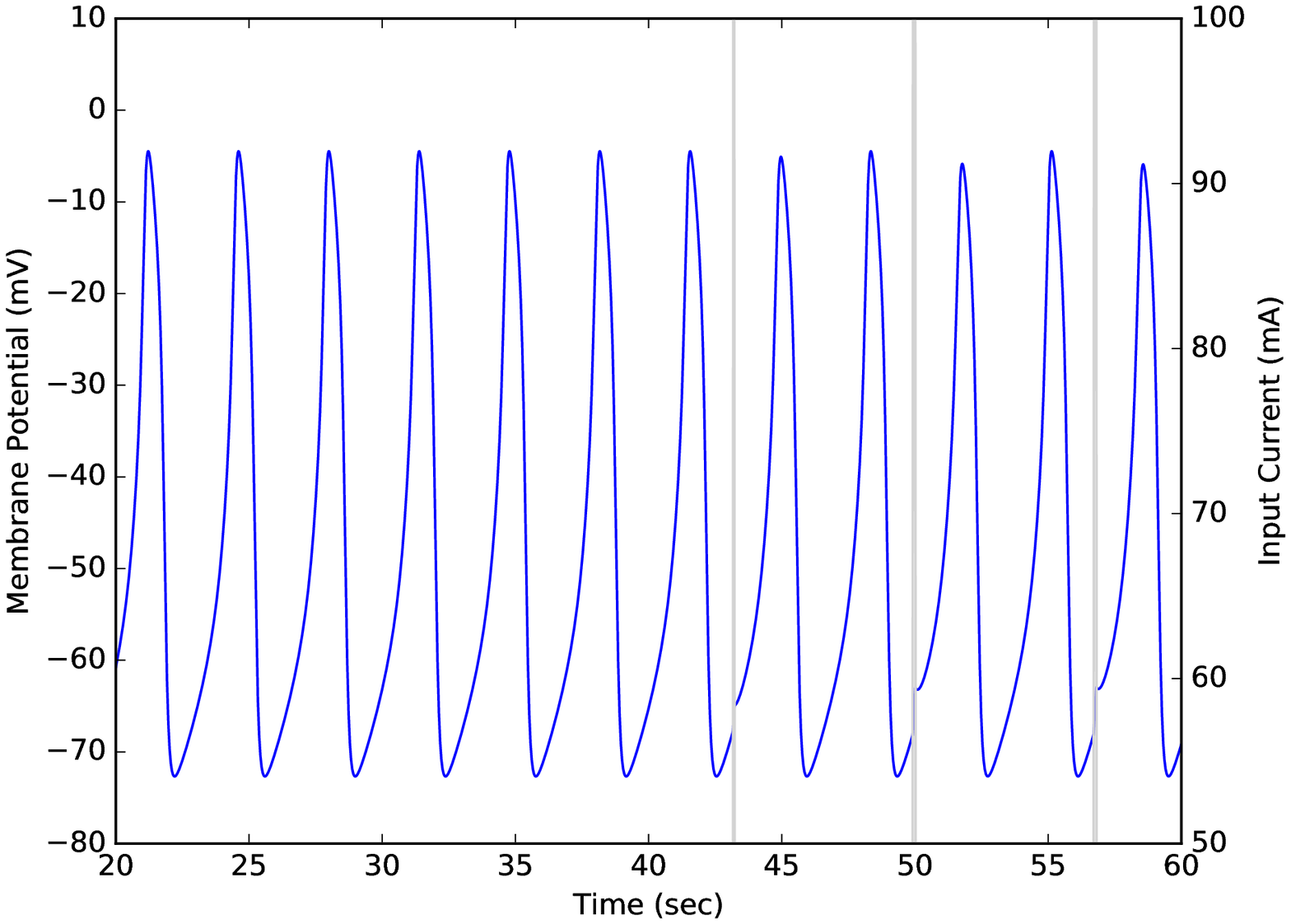}
\end{center}

\caption{Lack of synchronization of a single-neuron oscillator model to
rhythmic stimulation at the beginning of the stimulation period. Before
stimulation the model was running on a limit cycle at the second
harmonic of the stimulation frequency.}

\label{fig:2-1SyncNeuralModelBegining}
\end{figure}

\begin{figure}
\begin{center}
\includegraphics[width=4in]{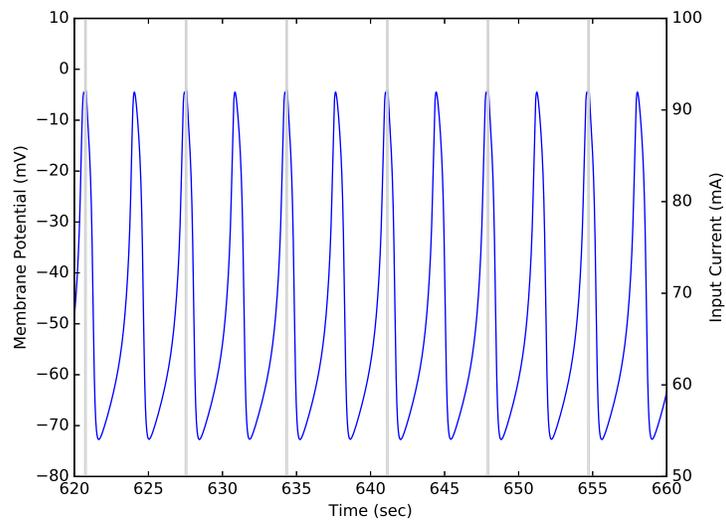}
\end{center}

\caption{Precise synchronization of a single-neuron oscillator model to
rhythmic stimulation by 640~seconds in the stimulation period. Before
stimulation the neural model was running on a limit cycle at the
second harmonic of the stimulation frequency.}

\label{fig:2-1SyncNeuralModelLater}
\end{figure}

\begin{figure}
\begin{center}
\includegraphics[width=4in]{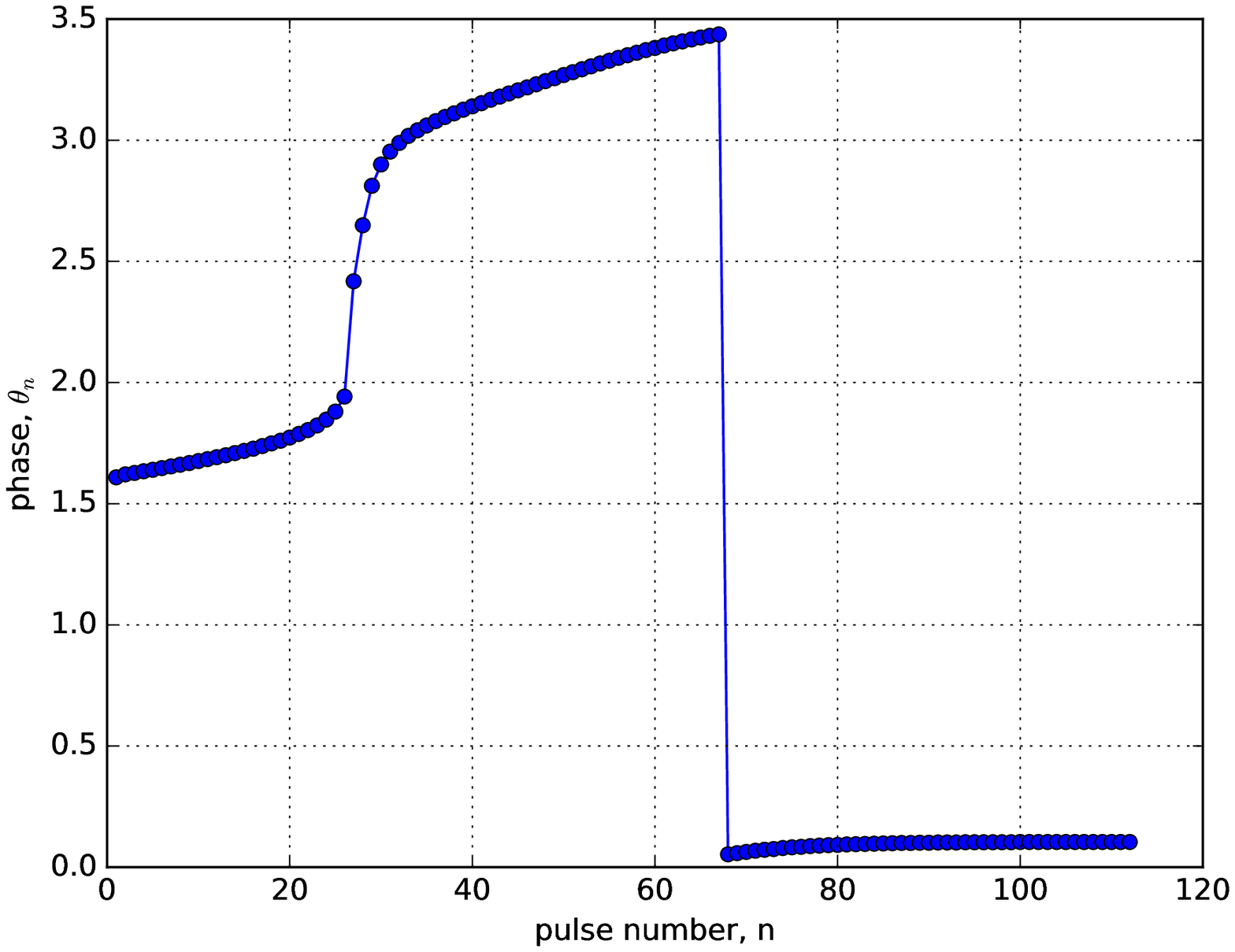}
\end{center}

\caption{Temporal evolution of synchronization for a single-neuron oscillator
model running on a limit cycle at the second harmonic of the
stimulation frequency. Blue points indicate phases of the spiking cycle at
which current input pulses were delivered to the model.}

\label{fig:2-1SyncNeuralModelAllTimes}
\end{figure}

Next we show that an oscillator running at the second harmonic of its input
frequency can also synchronize to its inputs
(see Faster Oscillator in Section~\ref{sec:oscillatorModels}).
Figures~\ref{fig:2-1SyncNeuralModelBegining}-\ref{fig:2-1SyncNeuralModelAllTimes}
are as
Figures~\ref{fig:1-1SyncNeuralModelBegining}-\ref{fig:1-1SyncNeuralModelAllTimes}
but for this faster oscillator. Figure~\ref{fig:2-1SyncNeuralModelLater} shows
that each other oscillation cycle is precisely synchronized with the inputs
(i.e., 2-1 synchronization). Figure~\ref{fig:2-1SyncNeuralModelAllTimes} again
shows that synchronization takes time to develop; for this faster oscillator
the establishment of synchronization took around 80 input pulses (544 seconds).

\subsection{Synchronization of neural populations to rhythmically produced CVSs}
\label{sec:syncInNeuralPopulations}

\begin{figure}
\begin{center}
\includegraphics[width=6in]{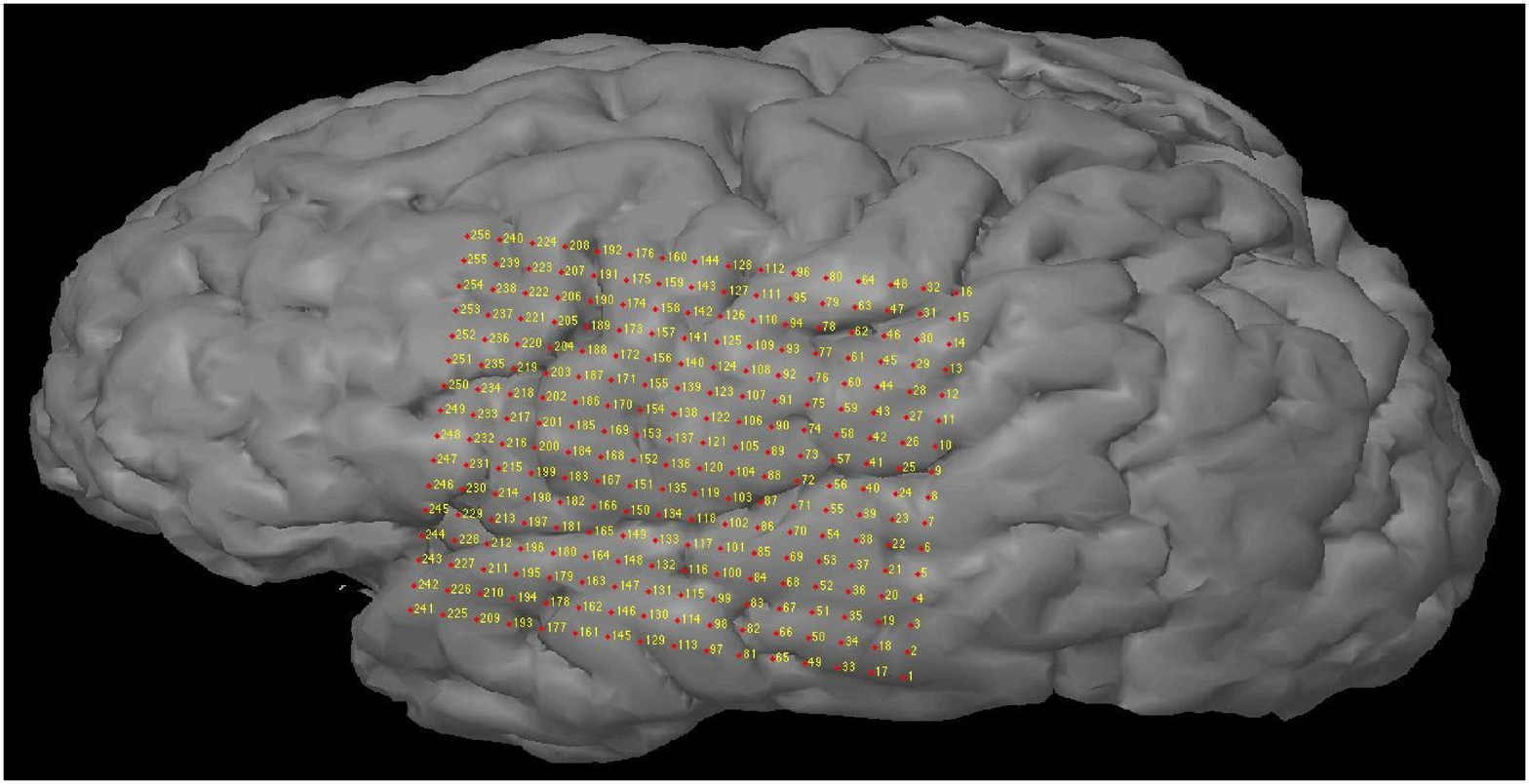}
\end{center}

\caption{Grid of ECoG electrodes superimposed on an MRI reconstruction of the
subjects' brain.}

\label{fig:grid}
\end{figure}

Here we demonstrate the synchronization of neural populations to rhythmically
produced \glspl{CVS} using \gls{ECoG} recordings from electrode 136
(Figure~\ref{fig:grid}) in a \gls{CVS}-production session.
Recording details are given in \citet{bouchardEtAl13}.
In Section~\ref{sec:syncTWs} we describe this
synchronization across all electrodes in the grid. We first study
synchronization in oscillators running at the median \gls{CVS}-production
frequency (fundamental frequency; Section~\ref{sec:1-1SyncECoG}), and then in
oscillators running at the twice this frequency (second-harmonic;
Section~\ref{sec:2-1SyncECoG}).

\subsubsection{1-to-1 synchronization at the fundamental frequency}
\label{sec:1-1SyncECoG}

\begin{figure}
\begin{center}
\includegraphics[width=3.5in]{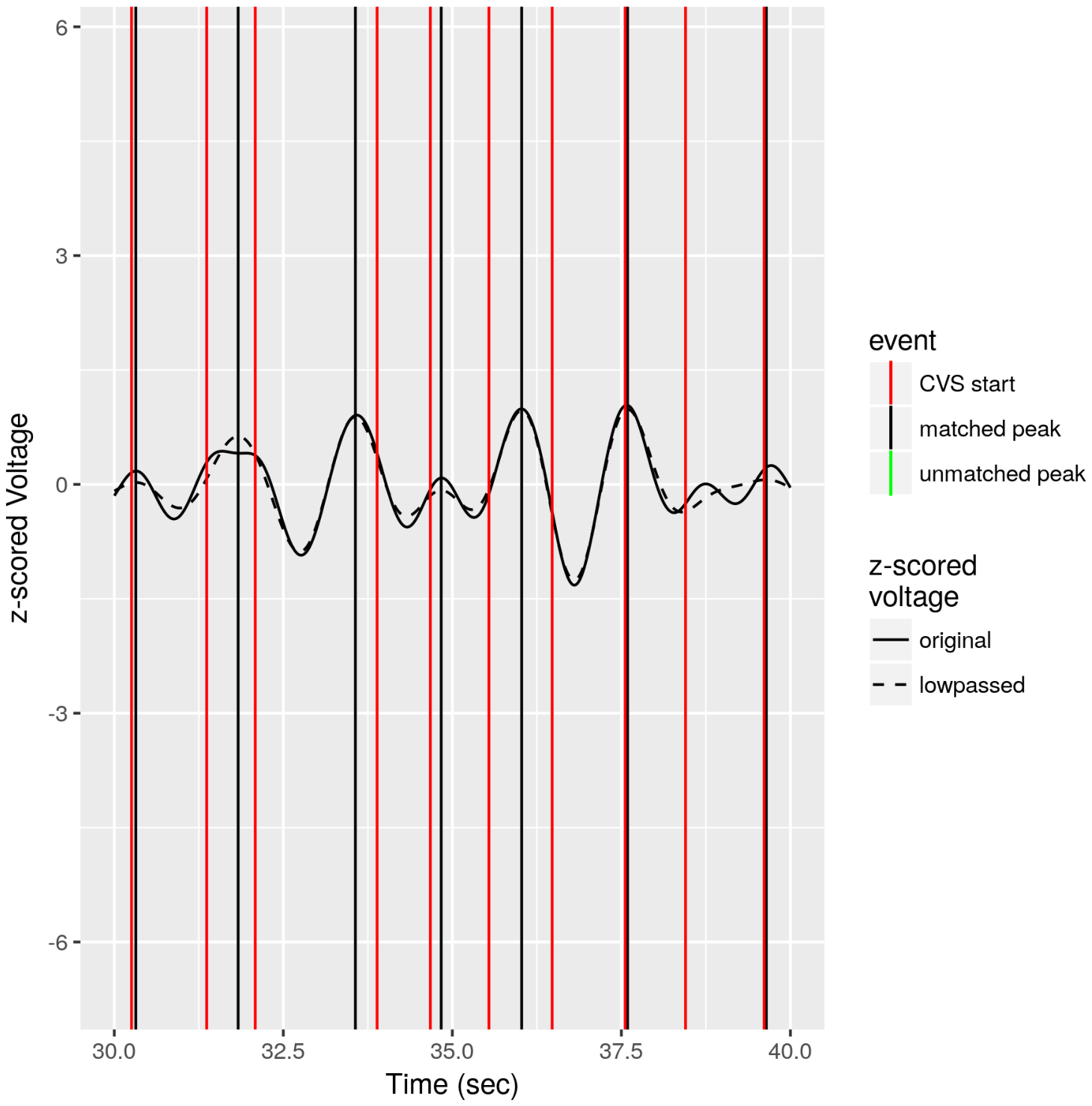}
\end{center}

\caption{At the beginning of the \gls{CVS}-production session, voltages from
electrode 136 filtered around the \gls{CVS}-production frequency are not
synchronized to \gls{CVS} initiations. The black trace plots the filtered
voltage, vertical black lines appear at peaks of the filtered voltage, and
vertical red lines indicate times of initiation of \gls{CVS} productions.}

\label{fig:1-1SyncECoGBegining}
\end{figure}

\begin{figure}
\begin{center}
\includegraphics[width=3.5in]{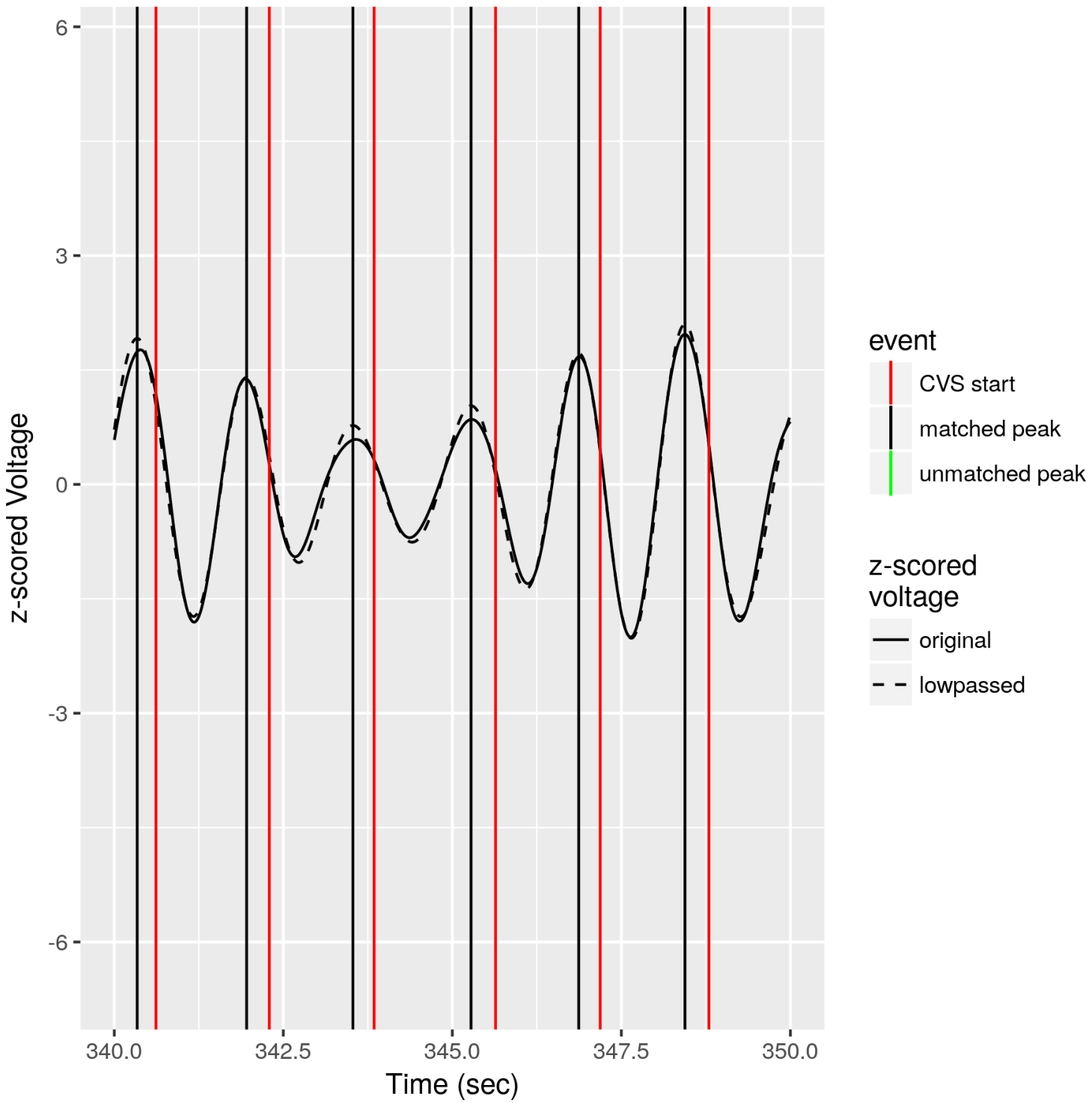}
\end{center}

\caption{Around 340~seconds into the \gls{CVS}-production session almost
perfect synchronization is achieved between \glspl{CVS} initiations and
voltages from electrode 136 filtered around the median \gls{CVS}-production
frequency. Same format as in Figure~\ref{fig:1-1SyncECoGBegining}.}

\label{fig:1-1SyncECoGLater}
\end{figure}

The black curve in Figures~\ref{fig:1-1SyncECoGBegining}
and~\ref{fig:1-1SyncECoGLater} plots voltages from electrode 136 bandpass
filtered around the median \gls{CVS}-production frequency at the beginning and
the middle of the \gls{CVS}-production session, respectively. Red vertical
lines mark times of initiation of \glspl{CVS} productions, and black vertical
lines mark peaks in voltage traces.
By the middle of the \gls{CVS}-production session
(Figure~\ref{fig:1-1SyncECoGLater}), but not at the beginning of this session
(Figure~\ref{fig:1-1SyncECoGBegining}), \glspl{CVS} initiations 
occur at precise times after filtered-voltage peaks (i.e.,
red vertical lines appear at similar delays after black vertical lines), and
filtered-voltage peaks and \gls{CVS} initiations are in a 1-to-1
correspondence (i.e., red vertical lines are matched to black vertical lines;
1-to-1 synchronization).

\begin{figure}
\begin{center}
\includegraphics[width=3.5in]{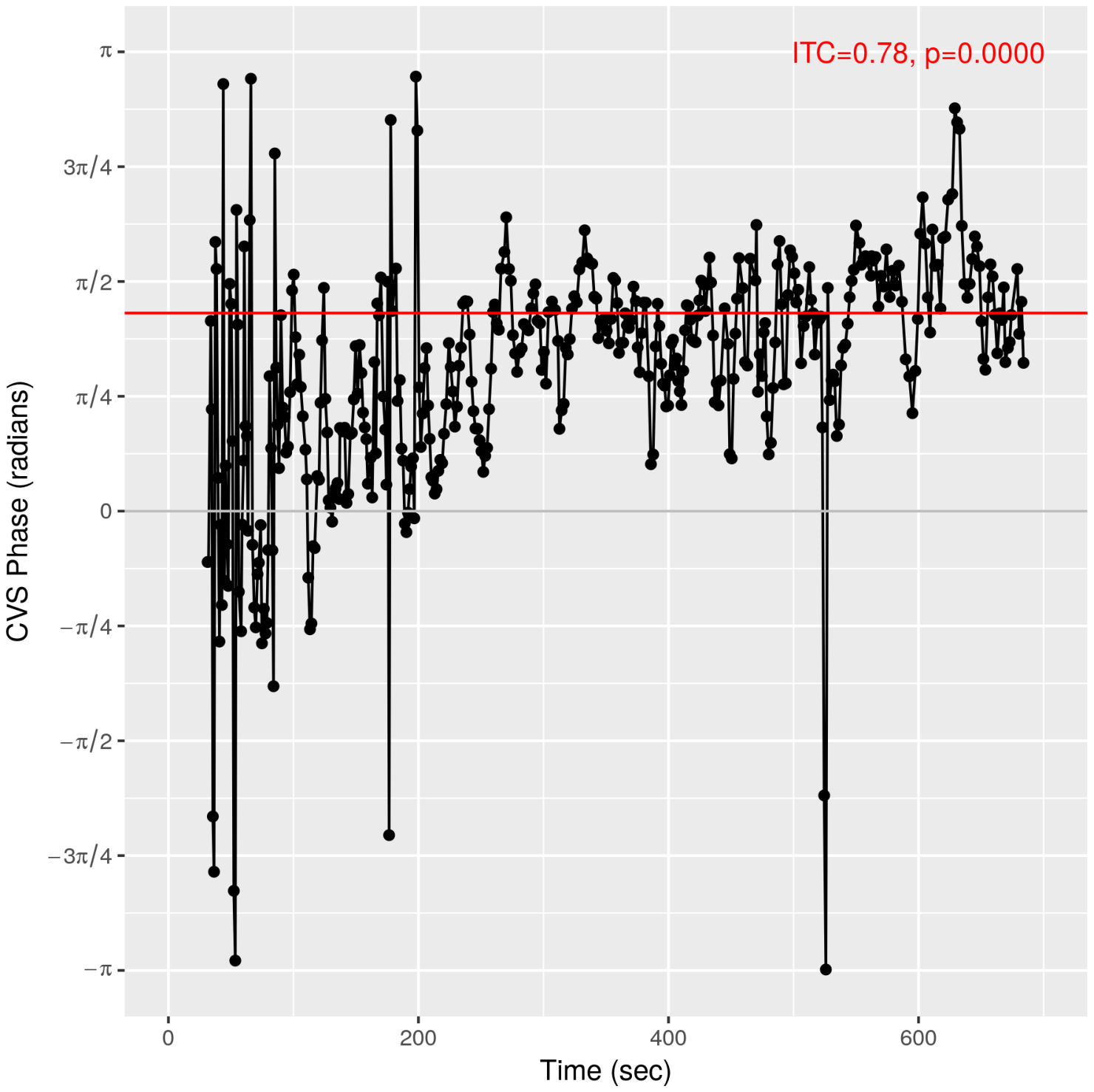}
\end{center}

\caption{Temporal evolution of synchronization between \gls{CVS} initiations
and voltages from electrode 136 filtered around the median \gls{CVS}-production
frequency.
Points indicate the phase of the filtered-voltage oscillation at which
\glspl{CVS} were initiated (i.e., \glspl{cvsPhase}). 
After an initial transient of around 200 seconds, \glspl{CVS} tend to be
initiated at a fixed phase of the filtered-voltage oscillations.
Between 400 seconds and the end of the \gls{CVS}-production session,
\glspl{cvsPhase} are highly concentrated (\gls{PLI}=0.78; $p<\num{1e-4}$,
Rayleigh non-uniformity test). The mean \gls{cvsPhase} (mean direction,
Section~\ref{sec:circularStats}) in this period is less than $\pi/2$ (red
horizontal line).
}

\label{fig:1-1CVSPhasesECoGAllTimes}
\end{figure}

For each \gls{CVS} production we measured the phase of the filtered
voltage at which the \gls{CVS} was initiated (i.e., the \gls{cvsPhase}).
Figure~\ref{fig:1-1CVSPhasesECoGAllTimes} plots these phases across the
\gls{CVS}-production session. After an initial transient period of around 200
seconds, \glspl{cvsPhase} become very precise.  To quantify this precision we
measured the Phase Locking Index (\gls{PLI}, Section~\ref{sec:pli}) of
these phases in the interval [400-700] seconds of the \gls{CVS}-production
session.  We obtained a \gls{PLI}=0.78, which was significantly different from
chance ($p<\num{1e-4}$, Rayleigh non-uniformity test). The mean phase (mean
direction, Section~\ref{sec:circularStats}) in this interval is indicated by
the red horizontal line in Figure~\ref{fig:1-1CVSPhasesECoGAllTimes}.

\begin{figure}
\begin{center}
\includegraphics[width=3.5in]{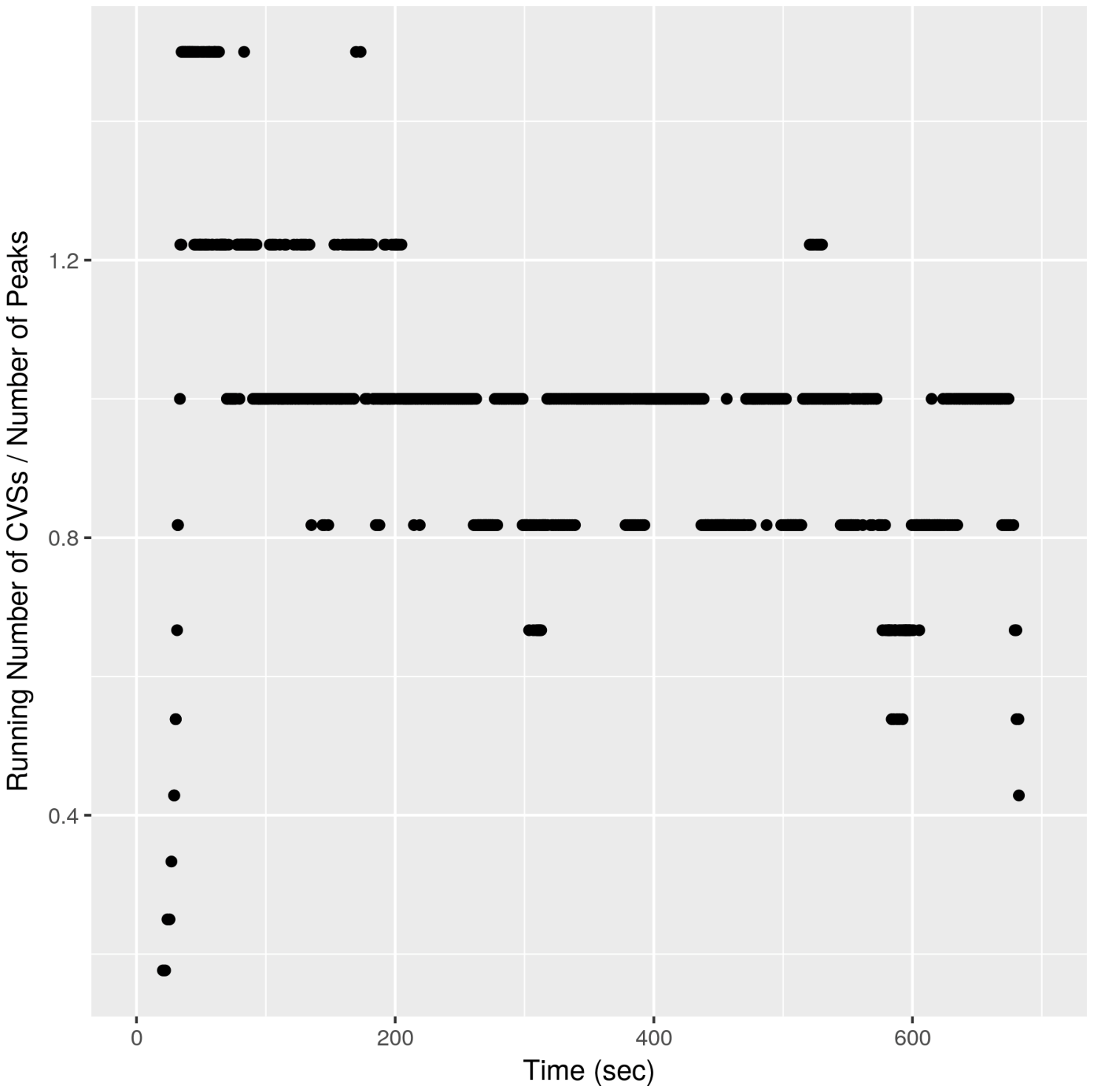}
\end{center}

\caption{After an initial transient period of around 200 seconds, a 1-to-1
synchronization developed between the initiations of \glspl{CVS} and filtered
voltages from electrode 136.
Voltages were filtered around the median \gls{CVS}-production frequency.
Points give the ratio between the number of \gls{CVS} initiations and the number of peaks
in the filtered voltage in a sliding window containing 20 \gls{CVS} initiations
and filtered-voltage peaks.
After the transient period, most ratios take the value 1.0, indicating that one
filtered-voltage peak is paired with one \gls{CVS} initiation (i.e, 1-to-1
synchronization).
}

\label{fig:1-1RatioNCVStoNPeaks}
\end{figure}

Figure~\ref{fig:1-1RatioNCVStoNPeaks} plots the running ratio of the number of
\gls{CVS} initiations to the number of filtered-voltage peaks in windows containing 20
\gls{CVS} initiations and filtered-voltage peaks. After an initial transient
period of around 200 seconds most windows show a ratio of 1.0, indicating that
\gls{CVS} initiations were paired with filtered-voltage peaks (i.e., 1-to-1
synchronization).

\subsubsection{2-to-1 synchronization at the second harmonic}
\label{sec:2-1SyncECoG}

\begin{figure}
\begin{center}
\includegraphics[width=3.5in]{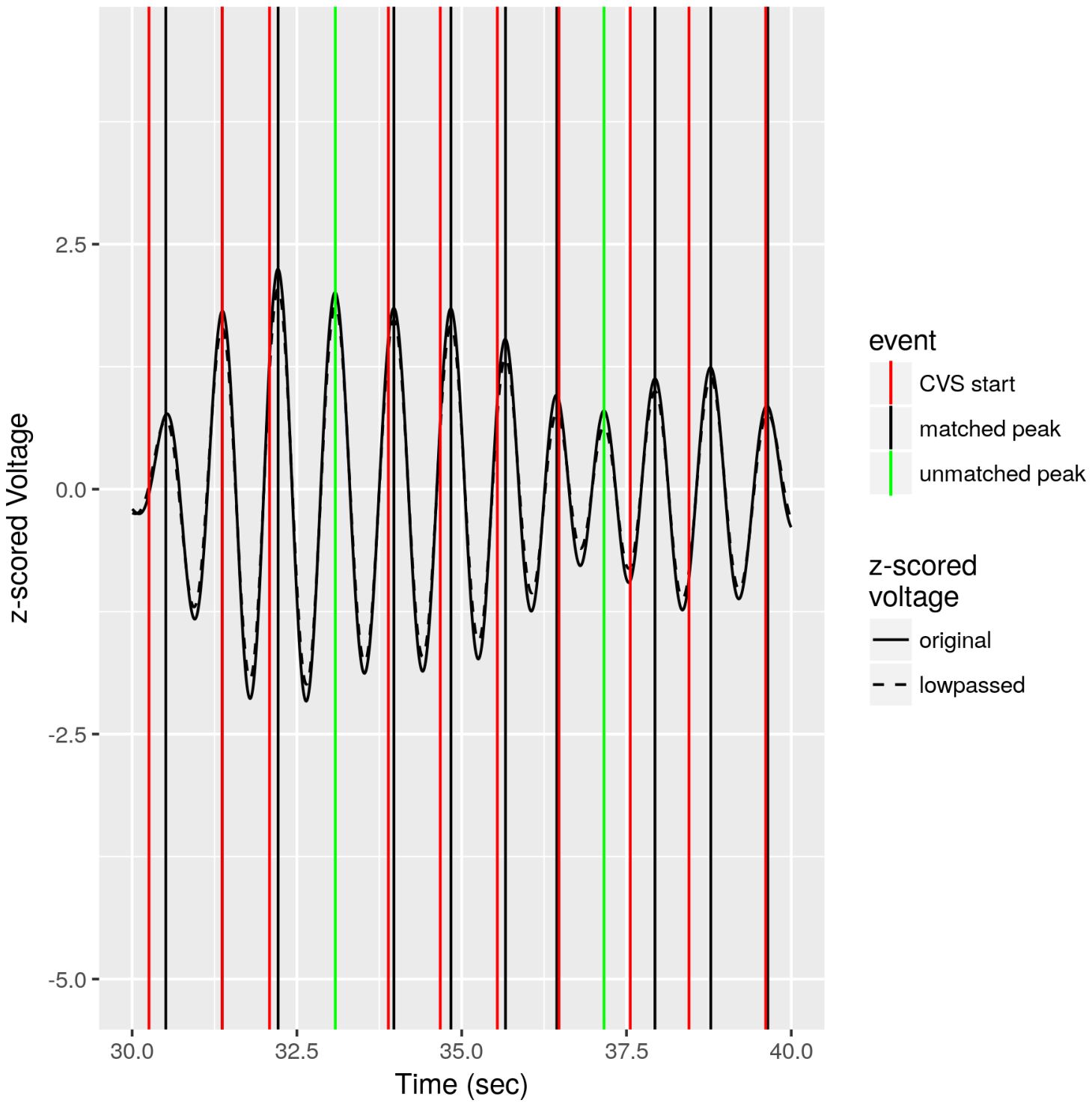}
\end{center}

\caption{At the beginning of the \gls{CVS}-production session no synchronization
is apparent between \gls{CVS} initiations and peaks of voltages from electrode 136 filtered around
the second harmonic of the \gls{CVS}-production frequency. Same format as in
Figure~\ref{fig:1-1SyncECoGBegining}.}

\label{fig:2-1SyncECoGBegining}
\end{figure}

\begin{figure}
\begin{center}
\includegraphics[width=3.5in]{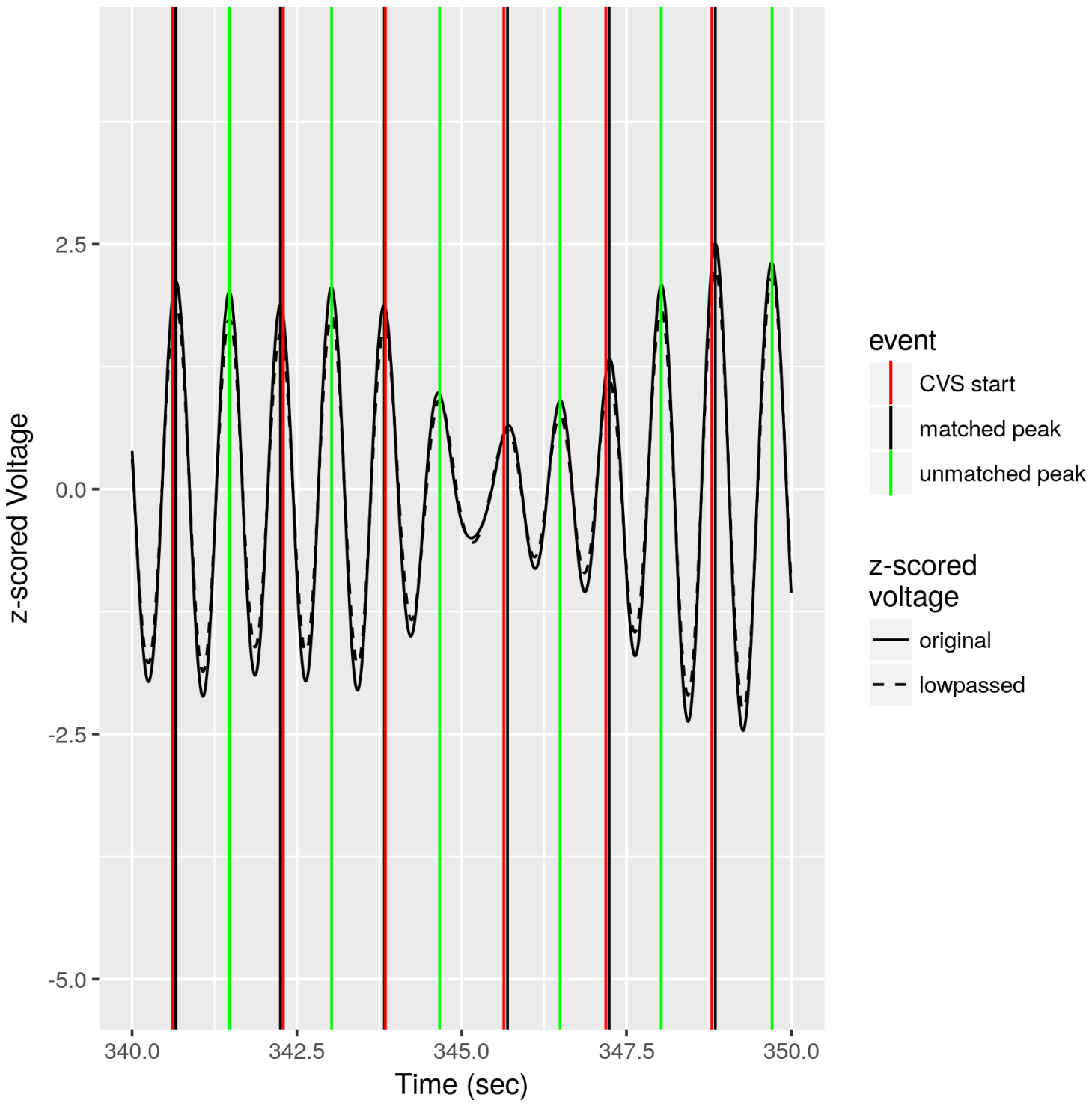}
\end{center}

\caption{By 340 seconds into the \gls{CVS}-production session an almost perfect
synchronization develops between the initiation of \glspl{CVS} and voltages
from electrode 136 filtered around the second harmonic of the median
\gls{CVS}-production frequency. The synchronization is 2-to-1 (two
filtered-voltage oscillations are paired with one \gls{CVS} initiation). Same
format as in Figure~\ref{fig:1-1SyncECoGBegining}. The green vertical lines
mark peaks of filtered voltages not paired to \gls{CVS} initiations.}

\label{fig:2-1SyncECoGLater}
\end{figure}

We bandpass filtered the
\gls{ECoG} recordings around the second \gls{CVS}-production frequency harmonic.
Figures~\ref{fig:2-1SyncECoGBegining} and~\ref{fig:2-1SyncECoGLater} are as
Figures~\ref{fig:1-1SyncECoGBegining} and~\ref{fig:1-1SyncECoGLater}, but for
voltages filtered around the second \gls{CVS}-production frequency harmonic.
Again we see that around 345~seconds in the \gls{CVS}-production session
(Figure~\ref{fig:2-1SyncECoGLater}), but not at the beginning of this session
(Figure~\ref{fig:2-1SyncECoGBegining}), the filtered-voltage
oscillations are precisely synchronized to the initiations of \gls{CVS}.
Similarly to
Figure~\ref{fig:2-1SyncNeuralModelLater}, Figure~\ref{fig:2-1SyncECoGLater}
shows that each other oscillation is synchronized to the initiation of a
\gls{CVS} (i.e., 2-to-1 synchronization).

\begin{figure}
\begin{center}
\includegraphics[width=3.5in]{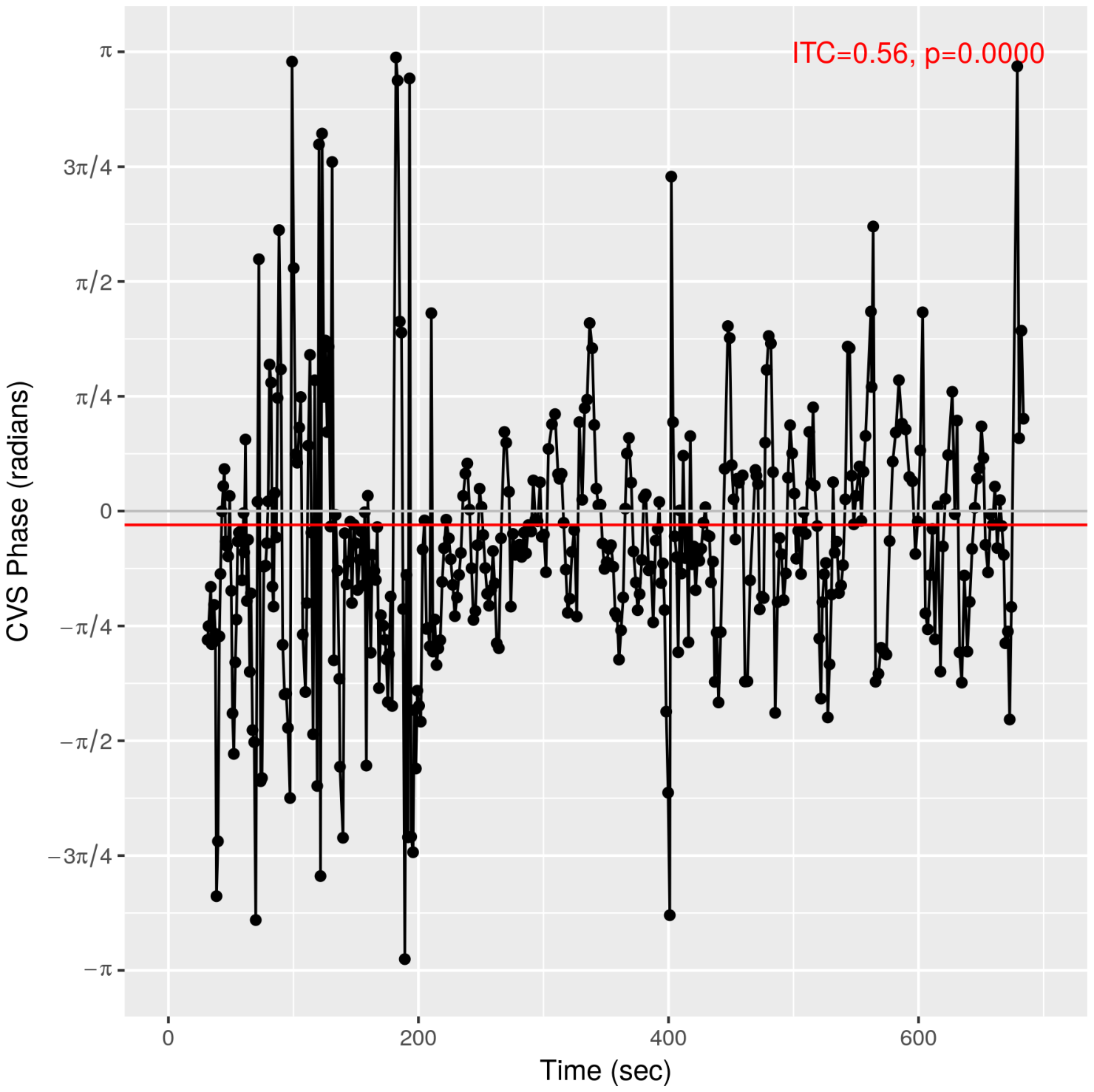}
\end{center}

\caption{Temporal evolution of synchronization between \gls{CVS} initiations
and voltage peaks from electrode 136 filtered around the second harmonic of the
\gls{CVS}-production frequency.
Same format as in Figure~\ref{fig:1-1CVSPhasesECoGAllTimes}.
After an initial transient period of around 200 seconds, \glspl{CVS} tend to
be initiated at peaks of the filtered-voltage oscillations (\gls{cvsPhase}=0).
Between 400 seconds and the end of the \gls{CVS}-production session,
\glspl{cvsPhase} are highly concentrated (\gls{PLI}=0.56; $p<\num{1e-4}$, Rayleigh
non-uniformity test) around the filtered-voltages peaks. The mean \gls{cvsPhase}
(mean direction, Section~\ref{sec:circularStats}) in this period is close to
zero (red horizontal line).}

\label{fig:2-1CVSPhasesECoGAllTimes}
\end{figure}

Figure~\ref{fig:2-1CVSPhasesECoGAllTimes} is as
Figure~\ref{fig:1-1CVSPhasesECoGAllTimes} but for voltages filtered around the
second harmonic of the \gls{CVS}-production frequency. \glspl{cvsPhase} were
significantly concentrated around zero (\gls{PLI}=0.56; $p<\num{1e-4}$ Rayleigh
non-uniformity test).

\begin{figure}
\begin{center}
\includegraphics[width=3.5in]{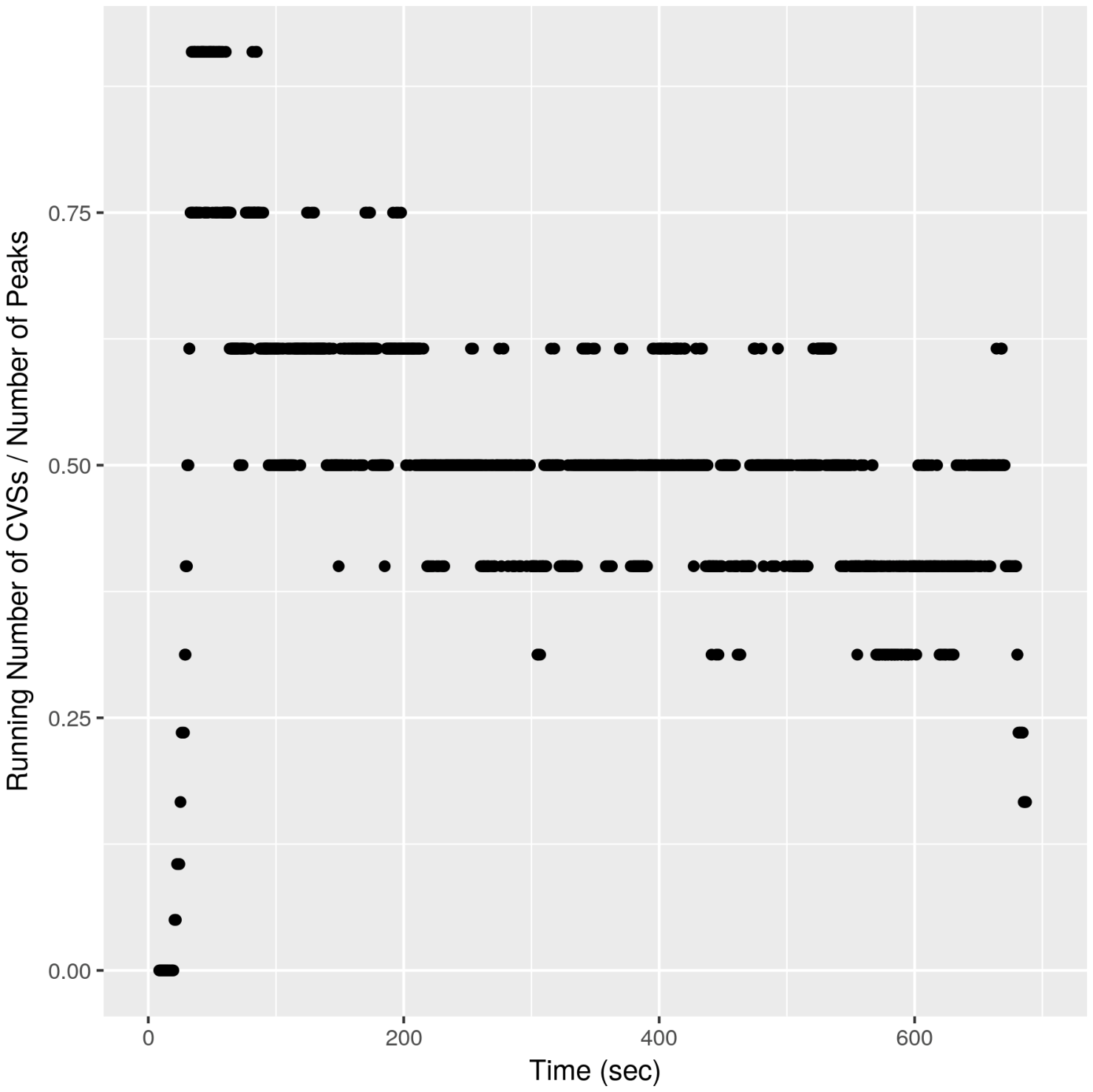}
\end{center}

\caption{After an initial transient period of around 200 second, a 2-to-1
synchronization developed between the initiation of \glspl{CVS} and voltages
filtered at the second harmonic of the \gls{CVS}-production frequency. Same format
as in Figure~\ref{fig:1-1RatioNCVStoNPeaks}. After the initial transient period
most ratios take the value 0.5, indicating that two filtered-voltage peaks
were paired with one \gls{CVS} initiation (i.e., 2-to-1 synchronization).}

\label{fig:2-1RatioNCVStoNPeaks}
\end{figure}

Figure~\ref{fig:2-1RatioNCVStoNPeaks} is as
Figure~\ref{fig:1-1RatioNCVStoNPeaks} but for voltages filtered around the
second harmonic of the \gls{CVS}-production frequency. After an initial
transient period of 200 seconds, for the majority of time windows, the ratio of
the number of \glspl{CVS} initiations to the number of filtered-voltage peaks
is 0.5, indicating that, as in Figure~\ref{fig:2-1SyncECoGLater}, each
\gls{CVS} initiation is paired to two peaks of filtered voltages (i.e., 2-to-1
synchronization).

\subsection{Synchronization of traveling waves to rhythmically produced
consonant-vowel syllables}
\label{sec:syncTWs}

The previous section showed that \glspl{cvsPhase} of voltages from electrode
136 were highly concentrated.
Section~\ref{sec:cvsPhaseaAcrossElectrodes} below demonstrates that this high
concentration extends across most electrodes in the grid, and shows that
\glspl{cvsPhase} are spatially organized consistently with the propagation of
\glspl{TW}.
Using this spatial organization, Section~\ref{sec:a1PremotorTWs} isolates an
extended \gls{TW} in voltages filtered around the median \gls{CVS}-production
frequency moving from primary auditory cortex to premotor cortex, and a
\gls{TW} in coupled high-gamma amplitude moving along the same path \gls{TW}
but in opposite direction.

\subsubsection{Synchronization of oscillators across electrodes}
\label{sec:cvsPhaseaAcrossElectrodes}

\begin{figure}
\begin{center}
\includegraphics[width=7.0in,angle=0]{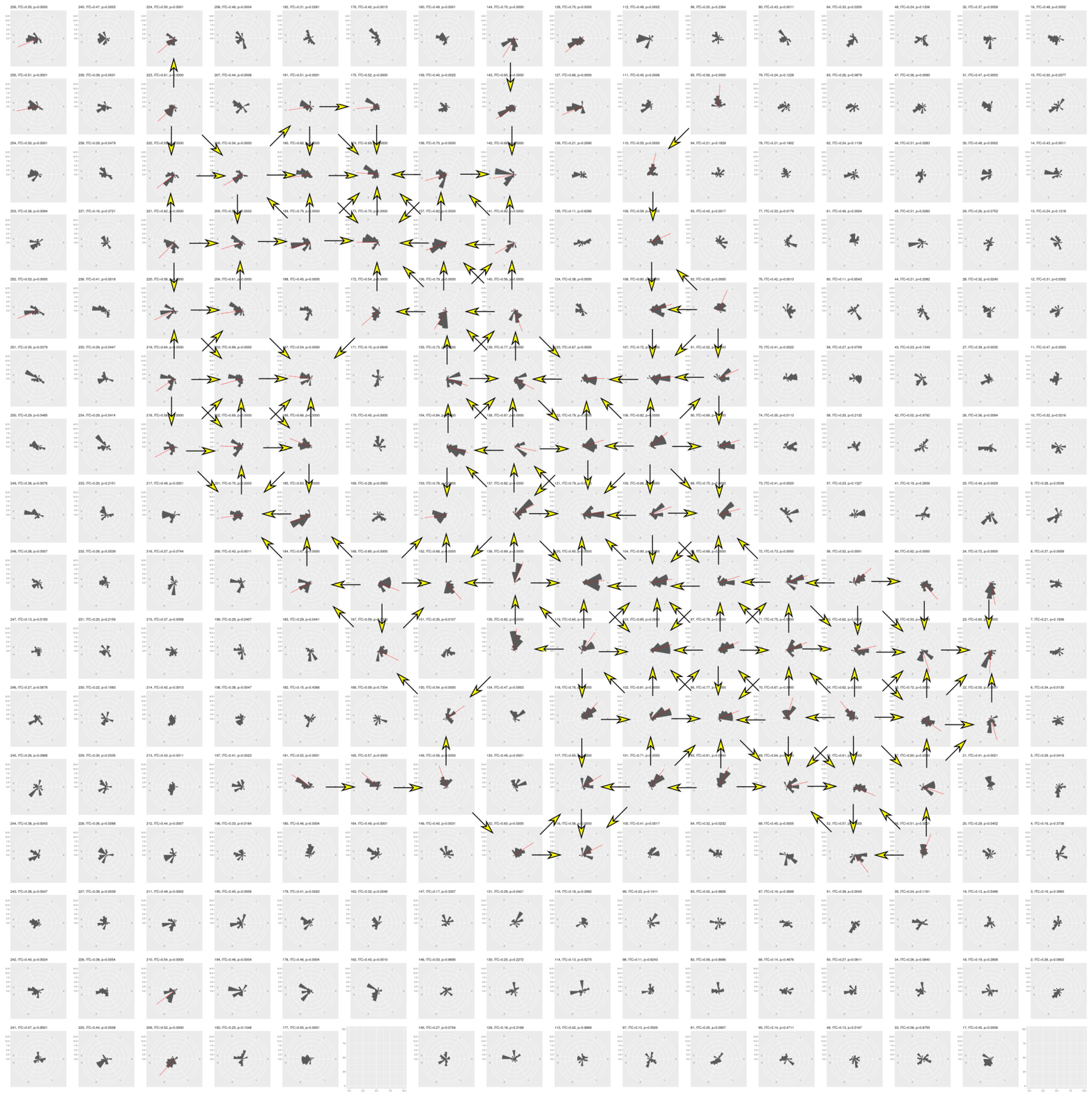}

\caption{
Histograms of \glspl{cvsPhase} for all electrodes in the grid.
Red segments are drawn at the mean \gls{cvsPhase} (mean direction,
Section~\ref{sec:circularStats}) in histograms with high \gls{cvsPhase}
concentration (\gls{PLI}\textgreater 0.5, Section~\ref{sec:pli}).
\glspl{cvsPhase} should decrease between one electrode and the next along the
direction of propagation of a \gls{TW} (i.e., red segments should move in the
clockwise direction).
Yellow arrows joining neighbor electrodes point in the direction consistent
with the propagation of a \gls{TW}.
The title of each histogram gives the electrode number, the \gls{PLI}, and the
p-value of a Rayleigh non-uniformity test. 
Histograms with high concentration of \glspl{cvsPhase} are mostly located in
the auditory cortex (bottom right), ventral sensorimotor cortex (center
middle), and premotor cortex (top left), and \glspl{TW} can be found between
some of these areas (e.g., Figure~\ref{fig:cvsPhasesAllElectrodesOneTW}).
}

\label{fig:cvsPhasesAllElectrodesManyTWs}
\end{center}
\end{figure}

Figure~\ref{fig:cvsPhasesAllElectrodesManyTWs} shows histograms of
\glspl{cvsPhase} for voltages filtered around the median \gls{CVS}-production
frequency between 340 and 400~seconds of the \gls{CVS}-production session.
In those histograms with high concentration of \glspl{cvsPhase} (i.e.,
\gls{PLI}\textgreater 0.5), a red segment indicates the mean phase (i.e., mean
direction, Section~\ref{sec:circularStats}).
In most
histograms the distribution of \glspl{cvsPhase} is highly concentrated
and
their mean phase orderly shifts from one electrode to its
neighbors (i.e., red segments change smoothly across the electrode grid).

The mean \gls{cvsPhase} should decrease as we move from one electrode
to the next along the propagation direction of a \gls{TW}. That is, in the
\gls{cvsPhase} histograms the red segment should move clockwise as we move
along the direction of propagation of a \gls{TW}. The yellow arrows in
Figure~\ref{fig:cvsPhasesAllElectrodesManyTWs} join neighboring histograms with
a high \gls{cvsPhase} concentration and
point in the direction consistent with the propagation of a \gls{TW}.

Histograms with a high \gls{cvsPhase} concentration are located in primary
auditory cortex (bottom right), in the ventral sensorimotor cortex (center
middle), and in the premotor cortex (top left) and \glspl{TW} can be found
between these areas, as illustrated next.

\subsubsection{TWs between premotor and primary auditory cortex synchronized to
the production of CVSs}
\label{sec:a1PremotorTWs}

\begin{figure}
\begin{center}
\includegraphics[width=7.0in,angle=0]{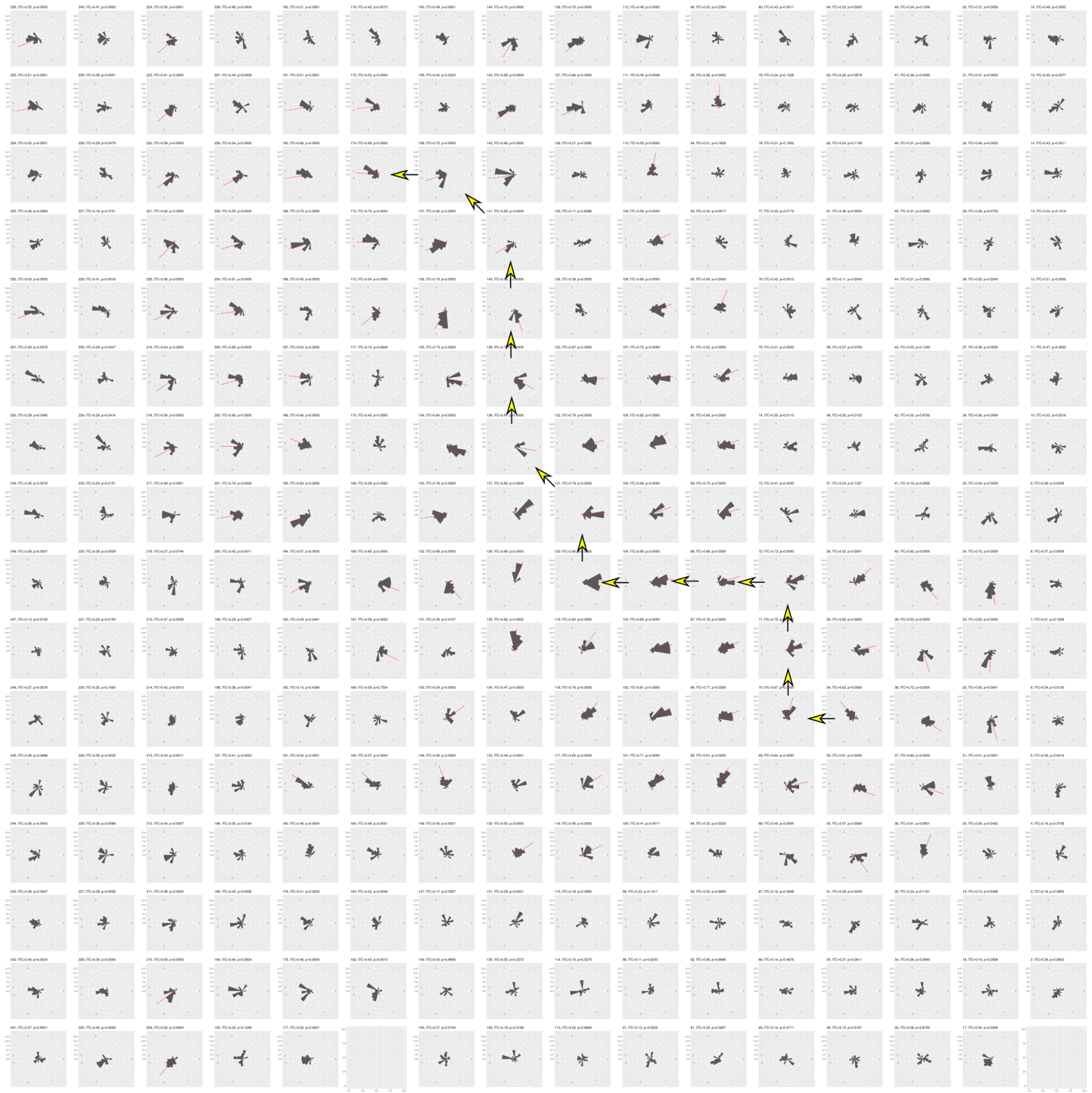}

\caption{
Histograms of \gls{CVS} phase gradients for all electrodes in the grid showing
a subset of the arrows in Figure~\ref{fig:cvsPhasesAllElectrodesManyTWs}
pointing in the direction of propagation of a \gls{TW} from primary auditory
cortex (electrode 54) to premotor cortex (electrode 174). Same format as in
Figure~\ref{fig:cvsPhasesAllElectrodesManyTWs}.
}

\label{fig:cvsPhasesAllElectrodesOneTW}
\end{center}
\end{figure}

Figure~\ref{fig:cvsPhasesAllElectrodesOneTW} depicts a subsets of the arrows in
Figure~\ref{fig:cvsPhasesAllElectrodesManyTWs} pointing along a path, from
primary auditory cortex (electrode 54) to premotor cortex (electrode 174), with
consistent direction of propagation of a \gls{TW}.  At the beginning of this
path in electrode 54 the mean \gls{cvsPhase} is close to $3/4\pi$.
Consistently with the propagation of a \gls{TW}, the mean \gls{cvsPhase}
decrease progressively along the path until electrode 174, where the mean
\gls{cvsPhase} is close to $-\pi$. That is, the wave traveled almost a full
cycle from auditory to premotor cortex.
The video at \url{https://youtu.be/CkQCau8RKvI} depicts voltages filtered
around the median \gls{CVS}-production frequency at the electrodes in this
path. A \gls{TW} moving from primary auditory to premotor cortex is evident.

Using the same recordings as those characterized here, in
\citet{rapelaInPrepTWsInSpeech} we reported \glspl{TW} in voltages filtered
around the median \gls{CVS}-production frequency and described a peculiar
organization of the coupling between phases of the filtered voltages and
high-gamma amplitudes.  We showed that this phase-amplitude coupling led to
\glspl{TW} of coupled high-gamma amplitude moving along the same path, but in
opposite direction, as \glspl{TW} of voltages filtered around the median
\gls{CVS}-production frequency.
The video at \url{https://youtu.be/8pdl6i7jnbw} shows a TW of coupled
high-gamma amplitude moving in the opposite direction as the \gls{TW} shown in
the previous video.
Both \glspl{TW} were extracted from a middle section of the \gls{CVS}-production
session (between 340 and 400~seconds from the start of the session).

It took several minutes from the start of the \gls{CVS}-production session for
the establishment of \glspl{TW}. The videos at
\url{https://youtu.be/NDrmVZ9lg9E} and at \url{https://youtu.be/n_4KWWrKfBQ}
show voltages filtered around the median \gls{CVS}-production frequency and
coupled high-gamma amplitudes, respectively, extracted from the initial section
of the \gls{CVS}-production section. \glspl{TW} in this initial section are
weaker than those in the middle section of this session.

\section{Discussion}

In \gls{ECoG} recordings from speech processing brain regions of a subject
rhythmically producing \glspl{CVS} here we reported that neural oscillators
were precisely synchronized to the production of \glspl{CVS}
(Section~\ref{sec:syncInNeuralPopulations}). We observed that the phase of the
oscillator cycle at wich \glspl{CVS} were initiated changed systematically
across the grid of electrodes, consistently with the propagation of \glspl{TW}
(Section~\ref{sec:cvsPhaseaAcrossElectrodes}). Using these synchronized phases
we isolated a first \gls{TW} in voltages filtered around the median
\gls{CVS}-production frequency moving from premotor to primary auditory cortex,
and a second \gls{TW} in high-gamma amplitude coupled to phase at the
\gls{CVS}-production frequency moving along the same path as the previous
\gls{TW} but in opposite riection (Section~\ref{sec:a1PremotorTWs}). These
\glspl{TW} may constitute a neural mechanisms for the interaction between the
production of speech and the perception of self-generated speech.

The observed synchronization of the initiation of \gls{CVS} productions and
\glspl{TW} is not a direct consequence of the rhythmic \gls{CVS}-production
task, since it took several minutes for the establishment of this
synchronization (Figures~\ref{fig:1-1CVSPhasesECoGAllTimes}
and~\ref{fig:2-1CVSPhasesECoGAllTimes}) and the syncrhonization was observed at
the second harmonic of the \gls{CVS}-production frequency
(Figures~\ref{fig:2-1SyncECoGLater} and~\ref{fig:2-1CVSPhasesECoGAllTimes}).

After an initial transient period of around 4 minutes, we observed precise
synchronization of oscillators across most electrodes of the array
(Figure~\ref{fig:cvsPhasesAllElectrodesManyTWs}). Future studies need to
investigate how this extended synchronization is achieved. Could it be the case
that exists an input (e.g., from Brocca's area) to a small group of oscillators
that synchronizes them to the production of \glspl{CVS}, and that
later the synchronization spreads to most electrodes in the array by weak
coupling between oscillators?

\section{Methods}

\subsection{Oscillator models}
\label{sec:oscillatorModels}

In Section~\ref{sec:syncInNeuralModels} we used the Persistent Sodium Potassium
model~\citep[INapIK; ][]{izhikevich07}.
The parameters of the slower oscillator in Section~\ref{sec:1-1SyncModels} were
as those in Figure~4.1a of \citet{izhikevich07} with a constant input current
$I=10$, and those of the faster oscillator in Section~\ref{sec:2-1SyncModels}
were as those in Figure~4.1b of \citet{izhikevich07} with a constant input
current $I=50$.

\subsection{Measuring CVS phases}
\label{sec:measuringCVSPhases}

A \gls{cvsPhase} measures in radians the location in an oscillatory cycle of
filtered voltages where the production of a \gls{CVS} is initiated. Call
$t_{CVS}$ the time of initiation of a \gls{CVS}, 
$t_{peak}$ the time of the peak of filtered voltages closest to $t_{CVS}$, 
$t_{next}$ the time of the first peak following the one at $t_{peak}$, and
$t_{prev}$ the time of the first peak preceding the one at $t_{peak}$.
We defined

\begin{eqnarray}
\text{CVS phase}(t_{CVS})=\frac{t_{CVS}-t_{peak}}{k}
\label{eq:cvsPhase}
\end{eqnarray}

\noindent If $t_{CVS}>t_{peak}$ then we used a normalization constant $k$
so that $\text{CVS Phase}(t_{next})=2\pi$ 
(i.e., $k=\frac{t_{next}-t_{peak}}{2\pi}$), and
if $t_{CVS}<t_{peak}$ then we took a normalization constant $k$
so that $\text{CVS phase}(t_{next})=-2\pi$
(i.e., $k=\frac{t_{peak}-t_{prev}}{2\pi}$).

\subsection{Circular statistics concepts}
\label{sec:circularStats}

This section introduces concepts from circular
statistics \citep{fisher96} used
to define \gls{PLI} in Section~\ref{sec:pli}.
Given a set of circular variables (e.g., phases), $\theta_1, \ldots,
\theta_N$, we associate to each circular variable a two-dimensional unit
vector. Using notation from complex numbers, the unit vector associated with
variable $\theta_i$ is:

\begin{eqnarray}
vec(\theta_i)=e^{j\theta_i}
\label{eq:unitVector}
\end{eqnarray}

\noindent The \emph{resultant vector}, $\mathbf{R}$, is the sum of the associated unit
vectors: 

\begin{eqnarray}
\mathbf{R}(\theta_1, \ldots, \theta_N)=\sum_{i=1}^{N}vec(\theta_i)
\label{eq:resultantVector}
\end{eqnarray}

\noindent The \emph{mean resultant length}, $\bar{R}$, is the length of the resultant
vector divided by the number of circular variables:

\begin{eqnarray}
\bar{R}(\theta_1, \ldots, \theta_N)=\frac{1}{N}|\mathbf{R}(\theta_1, \ldots,
\theta_N)|
\label{eq:meanResultantLength}
\end{eqnarray}

\noindent The \emph{circular variance}, $CV$, is one minus the mean resultant length:

\begin{eqnarray}
CV(\theta_1, \ldots, \theta_N)=1-\bar{R}(\theta_1, \ldots, \theta_N)
\label{eq:cv}
\end{eqnarray}

\noindent The \emph{mean direction}, $\bar{\theta}$, is the angle of the resultant
vector:

\begin{eqnarray}
\bar{\theta}(\theta_1, \ldots, \theta_N)=\arg(\mathbf{R}(\theta_1, \ldots,
\theta_N))
\label{eq:meanDirection}
\end{eqnarray}

\noindent Note that the mean direction is not defined when the resultant vector
is zero, since the angle of the zero vector is undefined.

\subsection{PLI}
\label{sec:pli}

The \gls{PLI}, also known as inter-trial coherence (ITC), is a measure of
coherence among a set of phases $\theta_1,\ldots,\theta_n$
\citep{tallonBaudryEtAl96, delormeAndMakeig04}.  It is the mean resultant
length ($\bar{R}$, Eq.~\ref{eq:meanResultantLength}) of these phases:

\begin{center}
\begin{eqnarray}
PLI(\theta_1,\ldots,\theta_N)=\bar{R}(\theta_1,\ldots,\theta_N)
\label{eq:pli}
\end{eqnarray}
\end{center}

\section{Acknowledgments}

We thank Dr.~Edward Chang and Dr.~Kristofer Bouchard for sharing the \gls{ECoG}
recordings.

\printglossary

\bibliographystyle{plainnatNoNote}
\bibliography{eeg,rhythms,speech,dynamicalSystems,synchronization,stats}

\begin{thebibliography}{37}
\providecommand{\natexlab}[1]{#1}
\providecommand{\url}[1]{\texttt{#1}}
\expandafter\ifx\csname urlstyle\endcsname\relax
  \providecommand{\doi}[1]{doi: #1}\else
  \providecommand{\doi}{doi: \begingroup \urlstyle{rm}\Url}\fi

\bibitem[Besle et~al.(2011)Besle, Schevon, Mehta, Lakatos, Goodman, McKhann,
  Emerson, and Schroeder]{besleEtAl11}
J.~Besle, C.A. Schevon, A.D. Mehta, P.~Lakatos, R.R. Goodman, G.M. McKhann,
  R.G. Emerson, and C.E. Schroeder.
\newblock Tuning of the human neocortex to the temporal dynamics of attended
  events.
\newblock \emph{The Journal of Neuroscience}, 31\penalty0 (9):\penalty0
  3176--3185, 2011.

\bibitem[Bouchard et~al.(2013)Bouchard, Mesgarani, Johnson, and
  Chang]{bouchardEtAl13}
K.E. Bouchard, N.~Mesgarani, K.~Johnson, and E.F. Chang.
\newblock Functional organization of human sensorimotor cortex for speech
  articulation.
\newblock \emph{Nature}, 495:\penalty0 327--332, 2013.

\bibitem[Cravo et~al.(2013)Cravo, Rohenkohl, Wyart, and Nobre]{cravoEtAl13}
A.M. Cravo, G.~Rohenkohl, V.~Wyart, and A.C. Nobre.
\newblock Temporal expectation enhances contrast sensitivity by phase
  entrainment of low-frequency oscillations in visual cortex.
\newblock \emph{The Journal of Neuroscience}, 33\penalty0 (9):\penalty0
  4002--4010, 2013.

\bibitem[Cummins(2012)]{cummins12}
F.~Cummins.
\newblock Oscillators and syllables: a cautionary note.
\newblock \emph{Frontiers in Psychology}, 3:\penalty0 364, 2012.

\bibitem[DeFelice and Isaac(1993)]{defeliceAndIsaac93}
Louis~J DeFelice and Aurora Isaac.
\newblock Chaotic states in a random world: Relationship between the nonlinear
  differential equations of excitability and the stochastic properties of ion
  channels.
\newblock \emph{Journal of Statistical Physics}, 70\penalty0 (1):\penalty0
  339--354, 1993.

\bibitem[Delorme and Makeig(2004)]{delormeAndMakeig04}
A.~Delorme and S.~Makeig.
\newblock {EEGLAB}: an open source toolbox for analysis of single-trial {EEG}
  dynamics including independent component analsys.
\newblock \emph{Journal of Neuroscience Methods}, 134\penalty0 (1):\penalty0
  9--21, 2004.

\bibitem[Ding and Simon(2014)]{dingAndSimon14}
N.~Ding and J.Z. Simon.
\newblock Cortical entrainment to continuous speech: functional roles and
  interpretations.
\newblock \emph{Front. Hum. Neurosci}, 8\penalty0 (311):\penalty0 10--3389,
  2014.

\bibitem[Doelling et~al.(2014)Doelling, Arnal, Ghitza, and
  Poeppel]{doellingEtAl14}
Keith~B Doelling, Luc~H Arnal, Oded Ghitza, and David Poeppel.
\newblock Acoustic landmarks drive delta--theta oscillations to enable speech
  comprehension by facilitating perceptual parsing.
\newblock \emph{Neuroimage}, 85:\penalty0 761--768, 2014.

\bibitem[Fisher(1996)]{fisher96}
N.I. Fisher.
\newblock \emph{Statistical analysis of circular data}.
\newblock Cambridge University Press, Cambridge, UK, 1996.

\bibitem[Galambos et~al.(1981)Galambos, Makeig, and Talmachoff]{galambosEtAl81}
Robert Galambos, Scott Makeig, and Peter~J Talmachoff.
\newblock A 40-hz auditory potential recorded from the human scalp.
\newblock \emph{Proceedings of the National Academy of Sciences}, 78\penalty0
  (4):\penalty0 2643--2647, 1981.

\bibitem[Glass(2001)]{glass01}
L.~Glass.
\newblock Synchronization and rhythmic processes in physiology.
\newblock \emph{Nature}, 410:\penalty0 227--284, 2001.

\bibitem[Golubitsky et~al.(1999)Golubitsky, Stewart, Buono, and
  Collins]{golubitskyEtAl99}
Martin Golubitsky, Ian Stewart, Pietro-Luciano Buono, and JJ~Collins.
\newblock Symmetry in locomotor central pattern generators and animal gaits.
\newblock \emph{Nature}, 401\penalty0 (6754):\penalty0 693--695, 1999.

\bibitem[Gomez-Ramirez et~al.(2011)Gomez-Ramirez, Kelly, Molholm, Sehatpour,
  Schwartz, and Foxe]{gomezRamirezEtAl11}
M.~Gomez-Ramirez, S.P. Kelly, S.~Molholm, P~Sehatpour, T.H. Schwartz, and J.J.
  Foxe.
\newblock Oscillatory sensory selection mechanism during intersensory attention
  to rhythmic auditory and visual inputs: a human electrocortigraphic
  investigation.
\newblock \emph{The Journal of Neuroscience}, 31\penalty0 (50):\penalty0
  18556--18567, 2011.

\bibitem[Graves et~al.(1986)Graves, Glass, Laporta, Meloche, and
  Grassino]{gravesEtAl86}
CARL Graves, LEON Glass, DONALD Laporta, ROGER Meloche, and ALEX Grassino.
\newblock Respiratory phase locking during mechanical ventilation in
  anesthetized human subjects.
\newblock \emph{American Journal of Physiology-Regulatory, Integrative and
  Comparative Physiology}, 250\penalty0 (5):\penalty0 R902--R909, 1986.

\bibitem[Gray et~al.(2015)Gray, Frey, Wilson, and Foxe]{grayEtAl15}
Michael~J Gray, Hans-Peter Frey, Tommy~J Wilson, and John~J Foxe.
\newblock Oscillatory recruitment of bilateral visual cortex during spatial
  attention to competing rhythmic inputs.
\newblock \emph{The Journal of Neuroscience}, 35\penalty0 (14):\penalty0
  5489--5503, 2015.

\bibitem[Gross et~al.(2013)Gross, Hoogenboom, Thut, Schyns, Panzeri, Belin, and
  Garrod]{grossEtAl13}
J.~Gross, N.~Hoogenboom, G.~Thut, P.~Schyns, S.~Panzeri, P.~Belin, and
  S.~Garrod.
\newblock Speech rhythms and multiplexed oscillatory sensory coding in the
  human brain.
\newblock \emph{PLoS Biology}, 11\penalty0 (12), 2013.
\newblock \doi{10.1371/journal.pbio.1001752}.

\bibitem[Guevara and Lewis(1995)]{guevaraEtAl95}
Michael~R Guevara and Timothy~J Lewis.
\newblock A minimal single-channel model for the regularity of beating in the
  sinoatrial node.
\newblock \emph{Chaos: An Interdisciplinary Journal of Nonlinear Science},
  5\penalty0 (1):\penalty0 174--183, 1995.

\bibitem[Izhikevich(2007)]{izhikevich07}
Eugene~M. Izhikevich.
\newblock \emph{Dynamical systems in neuroscience}.
\newblock MIT press, 2007.

\bibitem[Lakatos et~al.(2005)Lakatos, Shah, Knuth, Ulbert, Kamos, and
  Schroeder]{lakatosEtAl05}
P.~Lakatos, A.S. Shah, K.H. Knuth, I.~Ulbert, G.~Kamos, and C.E. Schroeder.
\newblock An oscillatory hierarchy controlling neural excitability and stimulus
  processing in the auditory cortex.
\newblock \emph{J. Neurophysiology}, 94\penalty0 (3):\penalty0 1904--1911,
  2005.

\bibitem[Lakatos et~al.(2008)Lakatos, Karmos, Mehta, Ulbert, and
  Schroeder]{lakatosEtAl08}
P.~Lakatos, G.~Karmos, A.D. Mehta, I.~Ulbert, and C.E. Schroeder.
\newblock Entrainment of neuronal oscillations as a mechanism of attentional
  selection.
\newblock \emph{Science}, 320:\penalty0 110--113, 2008.

\bibitem[Lakatos et~al.(2013)Lakatos, Musacchia, O'connel, Falchier, Javitt,
  and Schroeder]{lakatosEtAl13}
P.~Lakatos, G.~Musacchia, M.N. O'connel, A.Y. Falchier, D.C. Javitt, and C.E.
  Schroeder.
\newblock The spectrotemporal filter mechanism of auditory selective attention.
\newblock \emph{Neuron}, 77\penalty0 (4):\penalty0 750--761, 2013.

\bibitem[Mathewson et~al.(2010)Mathewson, Fabiani, Gratton, Beck, and
  Lleras]{mathewsonEtAl10}
K.E. Mathewson, M.~Fabiani, G.~Gratton, D.M. Beck, and A.~Lleras.
\newblock Rescuing stimuli from invisibility: Inducing a momentary release from
  visual masking with pre-target entrainment.
\newblock \emph{Cognition}, 115:\penalty0 186--191, 2010.

\bibitem[Millman et~al.(2015)Millman, Johnson, and Prendergast]{millmanEtAl15}
R.E. Millman, S.R. Johnson, and G.~Prendergast.
\newblock The role of phase-locking to the temporal envelope of speech in
  auditory perception and speech intelligibility.
\newblock \emph{Journal of Cognitive Neuroscience}, 2015.

\bibitem[O'Connell et~al.(2011)O'Connell, Falchier, McGinnis, Schroeder, and
  Lakatos]{oconnellEtAl11}
M.N. O'Connell, A.~Falchier, T.~McGinnis, C.E. Schroeder, and P.~Lakatos.
\newblock Dual mechanism of neuronal ensemble inhibition in primary auditory
  cortex.
\newblock \emph{Neuron}, 69\penalty0 (4):\penalty0 805--817, 2011.

\bibitem[Park et~al.(2015)Park, Ince, Schyns, Thut, and Gross]{parkEtAl15}
Hyojin Park, Robin~AA Ince, Philippe~G Schyns, Gregor Thut, and Joachim Gross.
\newblock Frontal top-down signals increase coupling of auditory low-frequency
  oscillations to continuous speech in human listeners.
\newblock \emph{Current Biology}, 2015.

\bibitem[Peelle and Davis(2012)]{peelleAndDavis12}
J.E. Peelle and M.H. Davis.
\newblock Neural oscillations carry speech rhythm through to comprehension.
\newblock \emph{Front Psychol}, 3\penalty0 (320):\penalty0 1--17, 2012.

\bibitem[Petrillo and Glass(1984)]{petrilloAndGlass84}
GA~Petrillo and LEON Glass.
\newblock A theory for phase locking of respiration in cats to a mechanical
  ventilator.
\newblock \emph{American Journal of Physiology-Regulatory, Integrative and
  Comparative Physiology}, 246\penalty0 (3):\penalty0 R311--R320, 1984.

\bibitem[Rapela(2016)]{rapelaInPrepTWsInSpeech}
Joaqu\'{i}n Rapela.
\newblock Entrainment of traveling waves to rhythmic motor acts, 2016.
\newblock URL \url{http://arxiv.org/abs/1606.02372}.

\bibitem[Regan(1966)]{regan66}
D~Regan.
\newblock Some characteristics of average steady-state and transient responses
  evoked by modulated light.
\newblock \emph{Electroencephalography and clinical neurophysiology},
  20\penalty0 (3):\penalty0 238--248, 1966.

\bibitem[Simon et~al.(2000)Simon, Habel, Daubenspeck, and Leiter]{simonEtAl00}
Peggy~M Simon, Alfred~M Habel, J~Andrew Daubenspeck, and JC~Leiter.
\newblock Vagal feedback in the entrainment of respiration to mechanical
  ventilation in sleeping humans.
\newblock \emph{Journal of Applied Physiology}, 89\penalty0 (2):\penalty0
  760--769, 2000.

\bibitem[Spaak et~al.(2014)Spaak, {de Lange}, and Jensen]{spaakEtAl14}
E.~Spaak, F.P. {de Lange}, and O~Jensen.
\newblock Local entrainment of alpha oscillations by visual stimuli causes
  cyclic modulation of perception.
\newblock \emph{J Neurosci}, 34\penalty0 (10):\penalty0 3536--44, 2014.

\bibitem[Strogatz(1986)]{strogatz86}
S.H. Strogatz.
\newblock \emph{The mathematical structure of the human sleep-wake cycle},
  volume~86.
\newblock Springer, 1986.

\bibitem[{Tallon Baudry} et~al.(1996){Tallon Baudry}, Bertrand, Delpuech, and
  Pernier]{tallonBaudryEtAl96}
C.~{Tallon Baudry}, O.~Bertrand, C.~Delpuech, and J.~Pernier.
\newblock Stimulus specificity of phase-locked and non-phase-locked 40 hz
  visual responses in human.
\newblock \emph{The Journal of Neuroscience}, 16:\penalty0 4240--4349, 1996.

\bibitem[Winfree(2001)]{winfree01}
Arthur~T Winfree.
\newblock \emph{The geometry of biological time}, volume~12.
\newblock Springer Science \& Business Media, 2001.

\bibitem[Zion~Golumbic et~al.(2012)Zion~Golumbic, Poeppel, and
  Schroeder]{zionGolumbicEtAl12}
E.M. Zion~Golumbic, D.~Poeppel, and C.E. Schroeder.
\newblock Temporal context in speech processing and attentional stream
  selection: a behavioral and neural perspective.
\newblock \emph{Brain and language}, 122\penalty0 (3):\penalty0 151--161, 2012.

\bibitem[{Zion Golumbic} et~al.(2013){Zion Golumbic}, Ding, Bickel, Lakatos,
  Schevon, McKhann, Goodman, Emerson, Mehta, Simon, Poeppel, and
  Schroeder]{zionGolumbicEtAl13}
E.M. {Zion Golumbic}, N.~Ding, S.~Bickel, P.~Lakatos, C.A. Schevon, G.M.
  McKhann, R.R. Goodman, R.~Emerson, A.D. Mehta, J.Z. Simon, D.~Poeppel, and
  C.E. Schroeder.
\newblock Mechanisms underlying selective neuronal tracking of attended speech
  at a ``cocktail party''.
\newblock \emph{Neuron}, 77\penalty0 (5):\penalty0 980--991, 2013.

\bibitem[Zion~Golumbic et~al.(2013)Zion~Golumbic, Ding, Bickel, Lakatos,
  Schevon, McKhann, Goodman, Emerson, Mehta, Simon, Poeppel, and
  Schroeder]{zionGolumbicEtAl13b}
E.M. Zion~Golumbic, N.~Ding, S.~Bickel, P.~Lakatos, C.A. Schevon, G.M. McKhann,
  R.R. Goodman, R.~Emerson, A.D. Mehta, J.Z. Simon, D.~Poeppel, and C.E.
  Schroeder.
\newblock Mechanisms underlying selective neuronal tracking of attended speech
  at a “cocktail party”.
\newblock \emph{Neuron}, 77\penalty0 (5):\penalty0 980--991, 2013.

\end{thebibliography}

\end{document}